\documentclass[preprint]{aastex}

\def	\be	{\begin{equation}}
\def	\ee	{\end{equation}}
\def	\beq	{\begin{eqnarray}}
\def	\eeq	{\end{eqnarray}}

\def	\simlt	{{\lower.5ex\hbox{$\;\buildrel{\mbox{\footnotesize$<$}}\over{\mbox{\footnotesize$\sim$}}\;$}}}
\def	\simgt	{{\lower.5ex\hbox{$\;\buildrel{\mbox{\footnotesize$>$}}\over{\mbox{\footnotesize$\sim$}}\;$}}}
\def	\define	{\stackrel{\rm def}{{}={}}}
\def	\etal	{{\em et~al.~}}

\def	\b		{{\hat b}}

\def	\k		{{\hat k}}
\def	\v		{{\mbox{\large$\upsilon$}}}
\def	\vT		{{\mbox{\large$\tilde\upsilon$}}}
\def	\vbf		{{\mbox{{\large\boldmath$\upsilon$}}}}

\def	\BST		{{\widetilde{B^2}}}

\def	\Bo		{{{}{}^0\!B}}
\def	\BoT		{{{}{}^0\!{\tilde B}}}
\def	\Bobf		{{\bf{}{}^0B}}
\def	\BoTbf		{{\bf{}{}^0{\tilde B}}}
\def	\bo		{{{}{}^0\b}}
\def	\bobf		{{\bf{}{}^0\b}}
\def	\vo		{{{}{}^0\v}}
\def	\vobf		{{\bf{}{}^0\vbf}}
\def	\bbo		{{{}{}^0b}}
\def	\bboT		{{{}{}^0{\tilde b}}}
\def	\MoT		{{{}{}^0\!{\cal M}}}		
\def	\Vo		{{{}{}^0V}}
\def	\BSo		{{{}{\vphantom{B^2}}^0\!B^2}}
\def	\BSoT		{{{}{\vphantom{B^2}}^0\!\BST}}

\def	\Bi		{{{}{}^1\!B}}
\def	\BiT		{{{}{}^1\!{\tilde B}}}
\def	\Bibf		{{\bf{}{}^1B}}
\def	\BiTbf		{{\bf{}{}^1{\tilde B}}}
\def	\bi		{{{}{}^1\b}}
\def	\bibf		{{\bf{}{}^1\b}}
\def	\vi		{{{}{}^1\v}}
\def	\viT		{{{}{}^1\vT}}
\def	\viTbf		{{\bf{}{}^1{\bf\tilde\vbf}}}
\def	\vibf		{{\bf{}{}^1\vbf}}
\def	\Vi		{{{}{}^1V}}
\def	\ViT		{{{}{}^1{\tilde V}}}
\def	\Vibf		{{\bf{}{}^1V}}
\def	\ViTbf		{{\bf{}{}^1{\tilde V}}}
\def	\bbi		{{{}{}^1b}}

\def	\Pi		{{{}{}^1\!P}}
\def	\PiT		{{{}{}^1\!{\tilde P}}}
\def	\FiT		{{{}{}^1\!{\cal F}}}		
\def	\BSi		{{{}{\vphantom{B^2}}^1\!B^2}}

\def	\Bii		{{{}{}^2\!B}}
\def	\Biibf		{{\bf{}{}^2B}}
\def	\BiiT		{{{}{}^2\!{\tilde B}}}
\def	\bii		{{{}{}^2\b}}
\def	\biibf		{{\bf{}{}^2\b}}
\def	\vii		{{{}{}^2\v}}
\def	\viiT		{{{}{}^2\vT}}
\def	\viibf		{{\bf{}{}^2\vbf}}
\def	\Vii		{{{}{}^2V}}
\def	\ViiT		{{{}{}^2{\tilde V}}}
\def	\Viibf		{{\bf{}{}^2V}}
\def	\ViiTbf		{{\bf{}{}^2{\tilde V}}}
\def	\Pii		{{{}{}^2\!P}}
		
\def	\BSii		{{{}{\vphantom{B^2}}^2\!B^2}}
\def	\BSiT		{{{}{\vphantom{B^2}}^1\!\BST}}
\def	\BSiiT		{{{}{\vphantom{B^2}}^2\!\BST}}

\received{2001 December 23}
\begin{document}

\title{Magnetized Turbulent Dynamo in Protogalaxies}

\author{Leonid Malyshkin}
\affil{The University of Chicago, ASCI Flash Center, 
Chicago, IL 60637-1433.\\ 
{\it leonmal@flash.uchicago.edu}}

\and

\author{Russell M. Kulsrud}
\affil{Princeton Plasma Physics Laboratory, Princeton, NJ 08543.\\ 
{\it rmk@pppl.gov}}

\date{\today}

\begin{abstract}
The prevailing theory for the origin of cosmic magnetic fields is that they have been 
amplified to their present values by the turbulent dynamo inductive action in the 
protogalactic and galactic medium. Up to now, in calculation of the turbulent dynamo, 
it has been customary to assume that there is no back reaction of the magnetic field 
on the turbulence, as long as the magnetic energy is less than the turbulent kinetic energy. 
This assumption leads to the kinematic dynamo theory. However, the applicability of this
theory to protogalaxies is rather limited. The reason is that in protogalaxies the 
temperature is very high, and the viscosity is dominated by magnetized ions. As the 
magnetic field strength grows in time, the ion cyclotron time becomes shorter than the ion 
collision time, and the plasma becomes strongly magnetized. As a result, the ion viscosity 
becomes the Braginskii viscosity. Thus, in protogalaxies the back reaction sets in much 
earlier, at field strengths much lower than those which correspond to field-turbulence 
energy equipartition, and the turbulent dynamo becomes what we call the magnetized 
turbulent dynamo. In this paper we lay the theoretical groundwork for the magnetized 
turbulent dynamo. In particular, we predict that the magnetic energy growth rate in the 
magnetized dynamo theory is up to ten time larger than that in the kinematic dynamo 
theory. We also briefly discuss how the Braginskii viscosity can aid the development of 
the inverse cascade of magnetic energy after the energy equipartition is reached. 
\end{abstract}


\section{Introduction}\label{INTRODUCTION}

One of the most important and challenging questions in astrophysics is the origin 
of strong and large-scale magnetic fields in galaxies and protogalaxies. It is now widely
accepted that the strong cosmic magnetic fields were produced by the turbulent dynamo 
inductive action driven by the fluid motions in a galactic and/or protogalactic medium
(Vainshtein \& Zel'dovich 1972; Zweibel \& Heiles 1997; Kulsrud 2000).
Yet full understanding of all stages of this production has not been achieved. 
There are two alternative theories on how and when the magnetic fields have been 
produced. The first theory, the galactic dynamo theory, also known as the $\alpha$--$\Omega$
dynamo theory, states that the fields have been primarily amplified in differentially 
rotating galactic disks after the galaxies had been formed (Parker 1971; Vainshtein \& 
Ruzmaikin 1972; Beck \etal 1996). The galactic dynamo involves several crucial 
unsolved problems (Rosner \& Deluca 1989; Zweibel \& Heiles 1997, Kulsrud 1999). 
The main problem is that in the $\alpha$--$\Omega$ theory it seems to be extremely 
difficult to expel a fraction of the magnetic flux from a galactic disk in order to 
produce the net magnetic flux (Rafikov \& Kulsrud 2000). In addition, observations
indicate the presence of microgauss magnetic fields in galaxy clusters and in early 
galaxies at high redshifts (Perry 1994; Kronberg 1994). It is hard to explain such 
strong fields by the galactic dynamo theory (Zweibel \& Heiles 1997). 
In this paper we accept the second theory for the origin of cosmic 
magnetic fields, {\it the primordial dynamo theory}, which states that the galactic 
and extragalactic magnetic fields have primarily been produced in protogalaxies, 
i.~e.~before the galaxies were formed (Pudritz \& Silk 1989; Kulsrud \& Anderson 1992; 
Kulsrud \etal 1997; Kulsrud 2000). Of course, these fields were subsequently 
modified in the rotating galactic disks after the galaxies were formed.

In order to understand how the magnetic fields can be built up in protogalaxies,
let us briefly discuss the physical conditions that were present 
there~\footnote{
The numerical quantities given below refer to common values at the time 
(redshift) when the protogalaxies form.
}. 
Let us assume the following typical values for the total mass $M$ of a 
protogalaxy, the total to baryon mass ratio $\xi$, and the protogalaxy size $L$: 
$M\sim 10^{12}\,{\rm M}_\odot$, $\xi\sim 10$, and $L\sim 0.2\,{\rm Mpc}$.
Then, the number density of the gas in the protogalaxy is $n\sim \xi^{-1}ML^{-3}m_p^{-1}
\sim 5\times 10^{-4}\,{\rm cm}^{-3}\propto \xi^{-1}ML^{-3}$ ($m_p$ is the proton 
mass, for convenience, in this paragraph we give the scaling of physical parameters 
with the three ``basic'' parameters $\xi$, $M$ and $L$). Assuming the energy virial 
equilibrium in the protogalaxy, we easily estimate the gas temperature, 
$T\sim k_B^{-1}Gm_pML^{-1}\sim 2\times 10^6\,{\rm K}\propto ML^{-1}$ ($G$ and 
$k_B$ are the gravitational and the Boltzmann constants). This temperature is very 
high, while the density is very low. As a result, the gas is fully ionized, 
and the viscosity is dominated by ions, not by neutrals. The ion collision time is 
$t_i\sim 20\,T^{3/2}n^{-1}\Lambda_c^{-1}\,{\rm sec}\sim 3\times 10^{12}\,{\rm sec}
\propto\xi M^{1/2}L^{3/2}$ (in this formula the temperature is in degrees K, 
the density is in ${\rm cm}^{-3}$, and $\Lambda_c\sim 30$ is the Coulomb logarithm 
assumed to be independent of $\xi$, $M$ and $L$, Braginskii 1965). The virial 
thermal speed is $V_T\sim(2k_BT/m_p)^{1/2}\sim 2\times 10^7\,{\rm cm}/{\rm s}\propto 
M^{1/2}L^{-1/2}$. The ion kinematic viscosity can be estimated as 
$\nu=0.96\,k_BTt_i/m_p\sim 5\times 10^{26}\,{\rm cm}^2/{\rm s}\propto\xi M^{3/2}L^{1/2}$
(Braginskii 1965). The Spitzer resistivity is $\eta_{\rm s}=
6.53\times 10^{12}\,T^{-3/2}\Lambda_c\,{\rm cm}^2/{\rm s}\sim 
8\times 10^4\,{\rm cm}^2/{\rm s}\propto M^{-3/2}L^{3/2}$ (in this formula the 
temperature is in degrees K, and the Coulomb logarithm is assumed to be constant, 
$\Lambda_c\sim 30$, Spitzer 1962). Now we estimate the Reynolds and the Prandtl 
numbers, $R\sim V_T/k_0\nu\sim 10^4\propto\xi^{-1}M^{-1}$ ($k_0=2\pi/L$ is the 
minimal wave number in the protogalaxy) and $Pr\sim \nu/\eta_{\rm s}\sim 
10^{22}\propto \xi M^3L^{-1}$, they are very large. The viscous cutoff scale of 
the turbulence can be estimated as $2\pi k_\nu^{-1}\sim R^{-3/4}L\sim 10^{-3}L\propto 
\xi^{3/4}M^{3/4}L$, the resistive cutoff scale for the magnetic field as 
$2\pi k_{\eta_s}^{-1}\sim Pr^{-1/2}R^{-3/4}L\sim 10^{-14}L\propto\xi^{1/4}M^{-3/4}L^{3/2}$,
and the ion mean free path as $\lambda_i\sim R^{-1}L\sim 10^{-4}L\propto \xi ML$.
Thus, there is a hierarchy of scales in the protogalaxy, $L \gg 2\pi k_\nu^{-1}\gg 
\lambda_i \gg 2\pi k_{\eta_s}^{-1}$. Therefore, we can use the single-fluid 
magnetohydrodynamic (MHD) equations for the description of plasma, and can consider 
the plasma to be nonresistive and incompressible on 
scales $L>2\pi k^{-1}\simgt\lambda_i$.~\footnote{
Note that the Debye length $(k_BT/2\pi ne^2)^{1/2}\sim 6\times 10^5\,{\rm cm}\propto
\xi^{1/2}L$ is extremely small in protogalaxies. Also we can consider 
the plasma to be incompressible on scales $<L$ because the plasma velocities at the 
largest scales, $\sim L$, are of the order of the sound speed, and all velocities at 
smaller scales are smaller. For estimation purposes, we can use the MHD equations 
even for scales $2\pi k^{-1}\ll\lambda_i$, if we reduce the molecular viscosity 
by factor $(k\lambda_i)^{-1}\ll 1$.
\label{COMMENT_MHD_TREATMENT}
}

It is very important that in a protogalaxy the ion cyclotron period in the 
magnetic field, $\omega_i^{-1}=m_pc/eB\sim 10^{-4}B^{-1}\,{\rm G}\cdot{\rm sec}$, 
is shorter than the ion collision time, $t_i\sim 3\times 10^{12}\,{\rm sec}$, 
provided that the field strength is larger than ${}\sim 10^{-17}$--$10^{-16}\,{\rm G}$. 
On the other hand, the magnetic energy becomes comparable to the kinetic energy of 
the smallest turbulent eddies (which are on the viscous cutoff scale) if the field 
strength exceeds ${}\sim (4\pi m_pnV_T^2R^{-1/2})^{1/2}\sim 10^{-7}\,{\rm G}$. 
Thus, there is a very broad range of magnetic field strengths, at which the magnetic 
pressure and tension are still negligible, while the presence of the field is 
already important. This is because the plasma is strongly magnetized, 
$\omega_i^{-1}\ll t_i$, and the magnetic field controls the microscopic motions of 
ions, so that the viscous forces are given by the Braginskii viscous stress 
[Braginskii 1965; see also eq.~(\ref{PI})] and are different from the viscous forces 
in a field-free plasma.

\clearpage

\begin{figure}[!t]
\vspace{7.0truecm}
\includegraphics{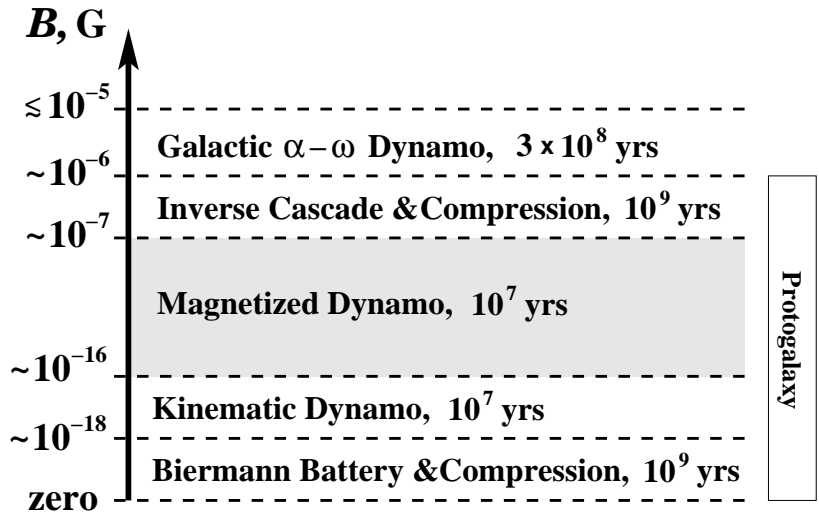}
\caption{
There are five major stages of the production of galactic and extragalactic 
magnetic fields, as the field strength grows up from zero to 
${}\simlt 10^{-5}\,{\rm gauss}$. In this paper we consider the magnetized 
turbulent dynamo stage in protogalaxies, shaded in the plot.
}
\label{FIGURE_FIELD_SCALES}
\end{figure}
\clearpage

We believe that there are five major stages of the production of the strong 
cosmic magnetic fields, see Figure~\ref{FIGURE_FIELD_SCALES}. 
During the first stage, in a protogalaxy undergoing gravitational collapse, the 
Biermann battery builds a seed magnetic field linearly in time, on a time scale 
approximately equal to the free-fall time, ${}\sim 1$~billion years. The 
resulting seed field is of the order of ${}\sim 10^{-18}\,{\rm G}$ on the 
viscous scale (Pudritz \& Silk 1989; Kulsrud \etal 1997; Davies \& Widrow 2000).

During the second stage, when the plasma is unmagnetized, $\omega_i^{-1}\gg t_i$, 
the seed field is exponentially amplified by the kinematic turbulent dynamo 
inductive action. The kinematic dynamo builds the field up to approximately 
$10^{-16}\,{\rm G}$, when the plasma becomes magnetized, $\omega_i^{-1}\ll t_i$. 
The time scale of the kinematic dynamo is very short, it is of the order of the 
smallest eddy turnover time (Kulsrud \etal 1997), ${}\sim 10$~million years, 
which is much smaller than the gravitational collapse time of the protogalaxy, 
several billion years. The main 
assumption of the kinematic dynamo theory and, basically, the strict definition of 
it, are that the growing magnetic field stays so weak, that it does not affect the 
fluid motions, i.~e.~that there is no {\it the back reaction} of the field on the 
turbulence. The results we obtain in this paper reduce to those of 
the kinematic dynamo theory in the limit of very weak field strengths (i.~e.~when 
the plasma is unmagnetized, and there is no the back reaction).

The third stage starts when the field grows above ${}\sim 10^{-16}\,{\rm G}$, the 
ion cyclotron time in the magnetic field becomes shorter than the ion collision 
time, and the plasma becomes strongly magnetized. During this stage the magnetic 
field strongly affects the dynamics of the turbulent motions on the viscous 
scales~\footnote{
The turbulent motions on the viscous scales are the most important 
in the dynamo theory. This is because the magnetic field is primarily amplified 
by the turbulent eddies on these scales. These eddies are the smallest ones,
they have the shortest turnover times and produce the largest velocity 
shearing.
\label{VISCOUS_IMPORTANT_SCALES}
},
by completely changing the viscosity (see Section~\ref{DYNAMOS}), despite the 
fact that the magnetic energy is still very small compared to the kinetic energy 
of the turbulence! We call this stage as {\it the magnetized turbulent dynamo}. 
In previous theories this stage has not been recognized. The main goal of this 
paper is to construct a theoretical model for it. The time scale of the magnetized 
dynamo is about the same as the kinematic dynamo time scale, ${}\sim 10$~million years.

So far the field scale is of the order of the viscous scale or less, and the 
magnetic field is incoherent in space. The fourth stage starts when the magnetic 
field strength grows above ${}\sim 10^{-7}\,{\rm G}$. The field energy becomes 
comparable to the kinetic energy of the smallest turbulent eddies, and the 
Lorentz forces become dynamically important in the plasma. 
During this stage the turbulent motions are dissipated by the growing field, the 
turbulent energy spectrum becomes truncated at larger and larger scales, and the 
turbulent energy is eventually transferred into a large-scale strong magnetic field 
with its energy comparable to the kinetic energy of the fluid motions on the largest 
scales in the protogalaxy (which is approximately the same as the thermal energy). 
This process (which is still under debate) is called the inverse cascade 
(Vainshtein 1982; Kulsrud \& Anderson 1992; Beck \etal 1996; Kulsrud 2000). We 
discuss it qualitatively in Section~\ref{DISCUSSION}. The time scale available 
for the inverse cascade process is of the order of the largest eddy turnover time, 
${}\sim 1$~billion years. 

The turbulent dynamos and the inverse cascade may not have time to amplify the field 
up to microgauss values, which are observed in galaxies. A crucial question is how 
far they go? This paper addresses this question and is concerned with the rate of 
the magnetic field built up by the magnetized turbulent dynamo. 

Finally, the fifth stage is the galactic dynamo, which happens in the differentially 
rotating galactic disc after the galaxy is formed. This process modifies the strong 
field that was initially built up in the protogalaxy, on a time scale of the order 
of the rotation time of the galaxy, ${}\sim 300$~million years. The galactic dynamo 
theory is not discussed in this paper.

In Section~\ref{DYNAMOS} we formulate the basic equations of the magnetized 
dynamo theory. In Section~\ref{VELOCITIES} we calculate the statistical correlation 
functions for turbulent velocities, ${\bf V}$, in a strongly magnetized plasma by 
making use of the quasilinear expansion in time of the MHD equations for both the 
velocities and the magnetic field, similar to the expansion used by Kulsrud 
and Anderson (1992). We find that, contrary to the Kolmogorov velocities, the 
turbulent velocities in the magnetized plasma are strongly anisotropic on 
the viscous scale, as one might expect, because the magnetic field sets 
``a preferred axis in space''. In our calculations we make two working hypotheses. 
First, we assume that for the purpose of magnetic energy calculation, the tensor 
$b_{\alpha\beta}=\b_\alpha\b_\beta$, where ${\bf\b}$ is the field unit vector, 
can be taken to be constant in space in the beginning of the expansion in time. 
This is our first hypothesis, which basically relies on our assumption that 
in the magnetized turbulent dynamo case the magnetic field has a folding 
structure similar to the one that exists in the kinematic turbulent dynamo case
(see Figure~\ref{FIGURE_FOLDING}; Maron \& Cowley 2001; Schekochihin \etal 2002). 
Second, we find that there are velocity modes which are not damped by the 
Braginskii viscous forces and, therefore, are divergent unless we incorporate 
the MHD non-linear inertial terms into our quasilinear expansion. We assume 
that these non-linear terms, which limit the divergent velocity modes by 
coupling them to other modes, may be included into our theory by allowing 
for rotation of velocity vectors relative to the magnetic field unit 
vectors~\footnote{
This rotation is essentially due to Coriolis forces that make the velocity rotate 
differently than the field direction.
}.
This is our second hypothesis.
In Section~\ref{MAGNETIC_SPECTRUM} we use the correlation functions for the 
turbulent velocities found in Section~\ref{VELOCITIES} to calculate the evolution
of the magnetic energy in the magnetized turbulent plasma. We start with 
calculations of the total magnetic energy growth rate in 
Subsection~\ref{MAGNETIC_ENERGY}. In Subsection~\ref{MODE_COUPLING_EQUATION} we 
derive the integro-differential mode coupling equation for the magnetic energy 
spectrum. Our mode coupling equation is a more general version of the 
corresponding equation of Kulsrud and Anderson (1992), obtained for the kinematic 
dynamo. In Subsection~\ref{SMALL_SCALES} we consider the magnetic energy spectrum 
on small subviscous scales. On these scales the mode coupling equation greatly 
simplifies and becomes a homogeneous differential equation. Finally, in 
Section~\ref{DISCUSSION} we give our conclusions. We also discuss the 
peculiarities of the inverse cascade in a strongly magnetized turbulent plasma.

\section
{Basic Magnetized Dynamo Equations}
\label{DYNAMOS}

Hereafter we consider the magnetized turbulent dynamo stage, the shaded region
in Figure~\ref{FIGURE_FIELD_SCALES}. During this stage the plasma is strongly 
magnetized, $\omega_i t_i\gg 1$, but the magnetic energy is still small 
compared to the turbulent kinetic energy, and therefore, the magnetic 
Lorentz forces can be neglected. The viscous forces acting on turbulent 
velocities ${\bf V}$ in a strongly magnetized incompressible fully 
ionized plasma are determined by the Braginskii viscosity stress tensor
(Braginskii 1965),
\beq
\pi_{\alpha\beta}=-\nu(3\b_\alpha\b_\beta-\delta_{\alpha\beta})
\b_\mu\b_\nu V_{\mu,\nu}\;,
\label{PI}
\eeq
where ${\bf\b}={\bf B}/B$ is the unit vector along the magnetic field. Note, 
that this stress tensor depends on the field unit vector, but is independent 
of the magnetic strength (as long as the plasma is strongly magnetized,
and $\omega_i t_i\gg 1$). Thus, during the magnetized turbulent dynamo 
stage the magnetic field strongly affects the turbulent motions on the viscous 
scales by changing the viscous forces, even though the Lorentz forces are 
negligible.

The MHD equations for the turbulent velocities ${\bf V}$ in an incompressible 
strongly magnetized plasma, neglecting the Lorentz forces, are 
(Landau \& Lifshitz 1984)
\beq
\partial_t V_\alpha &=&
-P'_{,\alpha}+f_\alpha-\pi_{\alpha\beta,\beta}-(V_\alpha V_\beta)_{,\beta}
\nonumber\\
{}&=&{}-P''_{,\alpha}+f_\alpha+3\nu(\b_\alpha\b_\beta\b_\mu\b_\nu V_{\mu,\nu})_{,\beta}
-(V_\alpha V_\beta)_{,\beta}\;,
\label{EQUATION_FOR_V}
\\
V_{\alpha,\alpha} &=& 0,
\label{DIV_V}
\eeq
where ${\bf f}$ is the force driving the turbulence, and $P'$ is the hydrodynamic 
pressure. Here and below we always assume summation over repeated indices. In 
order to shorten notations, we use $\partial_t\define\partial/\partial t$, 
and spatial derivatives are assumed to be taken with respect to all indices that 
are listed after ``$\,,\,$'' signs~\footnote{ 
For example,
$(V_\alpha V_\beta)_{,\gamma}\equiv V_\beta(\partial V_\alpha/\partial x_\gamma)+
V_\alpha(\partial V_{\beta}/\partial x_\gamma)$, and
$V_{\alpha,\beta\gamma}\equiv\partial^2V_{\alpha}/\partial x_\beta\partial x_\gamma$.
}.
To obtain the second line of equation~(\ref{EQUATION_FOR_V}), we use formula~(\ref{PI})
for the viscous stress and incorporate the isotropic part of the stress into the 
pressure $P''$.

It is difficult to solve equations~(\ref{EQUATION_FOR_V}) and~(\ref{DIV_V}) directly 
because they are very complicated. We also do not know the exact expression of the 
driving force ${\bf f}$, but its statistics is the same as for an unmagnetized plasma. 
Therefore, let us proceed as follows. We represent ${\bf f}$ by introducing subsidiary 
incompressible turbulent velocities ${\bf U}$ which, by definition, satisfy equations
\beq 
\partial_t U_\alpha &=&
-P'''_{,\alpha}+f_\alpha+(1/5)\nu\triangle U_\alpha-(U_\alpha U_\beta)_{,\beta}\;,
\label{EQUATION_FOR_U}
\\
U_{\alpha,\alpha} &=& 0,
\label{DIV_U}
\eeq
where $\triangle U_\alpha=U_{\alpha,\beta\beta}\,$. (${\bf U}$ will be essentially
the Kolmogorov turbulent velocities.)

Let us analyze and compare equations~(\ref{EQUATION_FOR_V}) and~(\ref{EQUATION_FOR_U}). 
First, if for the moment we formally average the Braginskii viscosity term 
$3\nu(\b_\alpha\b_\beta\b_\mu\b_\nu V_{\mu,\nu})_{,\beta}$ in 
equation~(\ref{EQUATION_FOR_V}) over all directions of an isotropic magnetic field,
then it reduces to $(1/5)\nu\triangle V_\alpha$, which coincides with the isotropic 
viscosity term in equation~(\ref{EQUATION_FOR_U}). Therefore, 
$\nu_{\rm eff}=(1/5)\nu$ could be considered as an {\it effective reduced viscosity}
for an incompressible fully ionized plasma in the presence of a magnetic field 
that is isotropically tangled on subviscous scales. In other words, the 
Braginskii viscous forces ``are doing worse'' at dissipating the turbulent 
motions, as compared with the standard isotropic viscous forces in a field-free 
plasma.
Second, note that equations~(\ref{EQUATION_FOR_V}) and~(\ref{EQUATION_FOR_U}) 
have the same driving force ${\bf f}$. By taking the driving force to be 
the same, we assume that this force comes from larger turbulent eddies. These 
larger eddies are on scales larger than the viscous scales, and therefore, 
these eddies ``do not know'' whether the viscous forces are of the Braginskii 
type or of the standard isotropic type.

Now, note that equation~(\ref{EQUATION_FOR_U}) is a familiar hydrodynamic 
equation with a standard isotropic viscosity term. However, is has a reduced 
molecular viscosity, $(1/5)\nu$ instead of $\nu$. Therefore, we assume that 
the solution of equations~(\ref{EQUATION_FOR_U}) and~(\ref{DIV_U}) is the 
incompressible, homogeneous, isotropic and stationary Kolmogorov turbulence 
with the effective reduced viscosity
\beq
\nu_{\rm eff}=(1/5)\nu.
\label{NU_EFF}
\eeq
As a result, the statistics of the Fourier coefficients of the 
turbulent velocities ${\bf U}$,
\beq
{\tilde U}_{{\bf k}\alpha}(\omega)= \frac{1}{\sqrt{2\pi}}\int_{-\infty}^\infty 
{\tilde U}_{{\bf k}\alpha}(t)\,e^{i\omega t}\,dt,
\qquad
{\tilde U}_{{\bf k}\alpha} (t)= \frac{1}{L^3}\int_{-L/2}^{L/2}
U_\alpha(t,{\bf r})\,e^{-i{\bf kr}}\,d^3{\bf r},
\label{FOURIER}
\eeq
is given by the following 
formulas~\footnote{
Note that these formulas are similar to those of Kulsrud \etal 1997, but differ 
by numerical coefficients from those of Kulsrud \& Anderson 1992.
}:
\beq
\langle {\tilde U}_{{\bf k}\alpha}(\omega)\rangle &=& 0, 
\label{U_STATISTICS}
\\
\langle {\tilde U}_{{\bf k}\alpha}(\omega){\tilde U}_{{\bf k'}\beta}(\omega')\rangle &=&
\langle {\tilde U}^*_{-{\bf k}\alpha}(-\omega){\tilde U}_{{\bf k'}\beta}(\omega')\rangle 
= J_{\omega k}\,\delta^\perp_{\alpha\beta}\,
\delta_{{\bf k'},{\bf -k}}\,\delta(\omega'+\omega), 
\qquad
\label{U_U_STATISTICS}
\\
\delta^\perp_{\alpha\beta} &\define& \delta_{\alpha\beta}-\k_\alpha\k_\beta,
\label{DELTA_PERP}
\\
J_{\omega k} &=& J_{0k}\left(1+\tau^2\omega^2\right)^{-1},
\label{J_OMEGA_K}
\\
\tau(k) &=& \tau(0)\,\left(k/k_0\right)^{-2/3}
=(1/k_0U_0)\left(k/k_0\right)^{-2/3},
\label{TAU}
\\
J_{0k} &\approx& 
\cases{
(U_0/6\pi k_0){(k/k_0)}^{-13/3},  &  $k_0\le k\le k_\nu=(5U_0/k_0\nu)^{3/4}k_0$, \cr
0,  &  $k<k_0=2\pi/L$,~~$k>k_\nu$.
}
\label{J_ZERO_K}
\eeq

Here and below $\langle ... \rangle$ means ensemble average over all realizations 
of the turbulence, $\delta(\omega'+\omega)$ is the Dirac $\delta$-function, 
$\delta_{\alpha\beta}$ and $\delta_{{\bf k'},{\bf -k}}$ are the one- and 
three-dimensional Kronecker symbols respectively, ${\bf\k}={\bf k}/k$ is the unit 
vector along $k$, $U_0\sim V_T$ is the largest eddy velocity, $k_0=2\pi/L$ is the 
smallest wave number of the turbulence, and 
$k_\nu\sim (U_0/k_0\nu_{\rm eff})^{3/4}k_0=(5U_0/k_0\nu)^{3/4}k_0$ is the viscous
cutoff wave number of the turbulence~\footnote{
$k_\nu$ is determined by the balance between the inertial and the viscous terms 
of eq.~(\ref{EQUATION_FOR_U}), $1/\tau(k_\nu)\sim\nu_{\rm eff}k_\nu^2$.
}. 
In equation~(\ref{U_U_STATISTICS}) we 
keep only the normal part of the turbulence and drop the helical part, since 
the latter is negligible on the scales of the smallest turbulent eddies, which 
are the principal drivers of the field evolution (Kulsrud \& Anderson 1992). 
To obtain equation~(\ref{J_OMEGA_K}), we assume that the time correlation 
function of the turbulent velocities has an exponential 
profile~\footnote{
Using a Gaussian time correlation profile, 
$\langle {\bf U}(t) {\bf U}(t')\rangle\propto e^{-(t-t')^2/2\tau^2}$, would be 
more appropriate. In this case equation~(\ref{J_OMEGA_K}) would become 
$J_{\omega k}=J_{0k}e^{-\tau^2\omega^2/2}$. However, we prefer the exponential 
profile because it is easier to deal with (e.g., for Gaussian integrals it is 
not possible to close integration contours at infinity in the complex plane).
}, 
\beq
\langle {\tilde U}_{{\bf k}\alpha}(t){\tilde U}_{{\bf k'}\beta}(t')\rangle &=&
\frac{J_{0k}}{2\tau} e^{-|t-t'|/\tau}\:\delta^\perp_{\alpha\beta}\,
\delta_{{\bf k'},{\bf -k}},
\label{U_U_TIME_CORRELATION}
\eeq
where $\tau$ is the eddy decorrelation time given by equation~(\ref{TAU})
for the Kolmogorov turbulence. Note that the averaged total kinetic energy of 
the fluid motions, per unit mass, is
\beq
\frac{1}{2}\langle[{\bf U}(t,{\bf r})]^2\rangle = 
\frac{1}{2}\sum_{\bf k} \langle|{\bf\tilde U}_{\bf k}(t)|^2\rangle = 
\frac{1}{2}\sum_{\bf k} \frac{J_{0k}}{\tau(k)} =
\frac{1}{2}\int_{k_0}^{k_\nu} I(k)\,dk=
\frac{1}{2}U_0^2,
\label{TOTAL_KINETIC_ENERGY}
\eeq
where the Kolmogorov energy spectrum is 
$I(k)=4\pi k^2 (L/2\pi)^3J_{0k}/\tau=(2/3)(U_0^2/k_0)(k/k_0)^{-5/3}$ if
$k\in[k_0,k_\nu]$, as it should be (Kulsrud \etal 1997).

Now let us subtract equations~(\ref{EQUATION_FOR_U}) and~(\ref{DIV_U}) from 
equations~(\ref{EQUATION_FOR_V}) and~(\ref{DIV_V}) to eliminate the unknown 
driving force ${\bf f}$, and let us introduce {\em the back-reaction velocity}~ 
$\vbf\define{\bf V}-{\bf U}$. We have
\beq
\partial_t \v_\alpha \!\!&=&\!\! {}-P_{,\alpha}
+3\nu(b_{\alpha\beta\mu\nu}\v_{\mu,\nu}+b_{\alpha\beta\mu\nu}U_{\mu,\nu})_{,\beta}
-(1/5)\nu\triangle U_\alpha
-(\v_\alpha U_\beta+U_\alpha\v_\beta+\v_\alpha\v_\beta)_{,\beta}\,,
\qquad
\label{v_EVOLUTION}
\\
\v_{\alpha,\alpha} \!\!&=&\!\! 0,
\label{v_DIV}
\eeq
where the pressure $P=P''-P'''$. Here and below we use the following symmetric 
tensors:
\beq
b_{\alpha\beta\gamma\delta} \define \b_\alpha\b_\beta\b_\gamma\b_\delta\;,
\qquad
b_{\alpha\beta\gamma} \define \b_\alpha\b_\beta\b_\gamma\;,
\qquad
b_{\alpha\beta} \define \b_\alpha\b_\beta\;.
\label{bb}
\eeq
Velocity $\vbf$, which satisfies equations~(\ref{v_EVOLUTION}) 
and~(\ref{v_DIV}), can be considered as the correction to the Kolmogorov 
velocity ${\bf U}$. This correction is non-zero only on the viscous scales,
and results from the strong back reaction of the field on the turbulence 
via the Braginskii viscosity tensor~(\ref{PI}). 

Finally, the MHD equation for the magnetic field ${\bf B}$ is 
(Landau \& Lifshitz 1984)
\beq
\partial_t B_\alpha &=& 
V_{\alpha,\beta} B_\beta - V_\beta B_{\alpha,\beta}\;,
\label{B_EVOLUTION}
\eeq
where the plasma velocities ${\bf V}={\bf U}+\vbf$ are incompressible, and 
we neglect resistivity. Consequently, the equations for the magnetic field 
squared, $B^2$, and for the magnetic field unit vector, ${\bf\b}$, are
\beq
\partial_t B^2 &=& 
2V_{\alpha,\beta} B_\alpha B_\beta - V_\beta (B^2)_{,\beta}\;,
\label{B_STRENGTH_EVOLUTION}
\\
\partial_t\,\b_\alpha &=& V_{\alpha,\beta}\b_\beta
-V_{\beta,\gamma}b_{\alpha\beta\gamma}-V_\beta\b_{\alpha,\beta}\;.
\label{b_EVOLUTION}
\eeq

\section
{Statistics of Turbulent Velocities in Strongly Magnetized Plasmas}
\label{VELOCITIES}

In order to find the evolution of the magnetic field ${\bf B}$, we must
derive the correlation functions for the total velocities ${\bf V}$, resulting
from the Braginskii viscosity. We calculate these velocity correlation 
functions in this section.

Let us assume that we know the magnetic field at zero time, 
${{\bf B}|}_{t=0}=\Bobf({\bf r})$ and ${{\bf\b}|}_{t=0}=\bobf({\bf r})$, 
and that the back-reaction velocity $\vbf$ is initially zero, 
${\vbf|}_{t=0}=\vobf({\bf r})\equiv 0$.~\footnote{
A nonzero initial back-reaction velocity would lead to transients, which would be 
dissipated anyway.
}
Then we advance the magnetic field and the back-reaction velocity to some future 
time, $t>0$, by the nonlinear terms, i.~e.~by integrating 
equations~(\ref{v_EVOLUTION}),~(\ref{B_EVOLUTION}) and~(\ref{b_EVOLUTION})
twice in time. This quasi-linear expansion procedure is similar to the calculations 
of Kulsrud and Anderson (1992). Considering $t$ as the expansion 
parameter~\footnote{
To be more formal, we need to introduce a dimensionless variable $\xi=t/\Delta t$, 
and to consider $U_{\alpha,\beta}\Delta t$ and $V_{\alpha,\beta}\Delta t$, which 
are dimensionless, as the expansion parameters.
}, 
up to the second order, we have
\beq
{\bf B}(t,{\bf r}) &=& \Bobf({\bf r})+\Bibf(t,{\bf r})+\Biibf(t,{\bf r}),
\label{B_EXPANSION}
\\
{\bf\b}(t,{\bf r}) &=& \bobf({\bf r})+\bibf(t,{\bf r})+\biibf(t,{\bf r}),
\label{b_EXPANSION}
\\
\vbf(t,{\bf r}) &=& \vibf(t,{\bf r})+\viibf(t,{\bf r}),
\label{v_EXPANSION}
\\
{\bf V}(t,{\bf r}) &=& \Vibf(t,{\bf r})+\Viibf(t,{\bf r}) 
= \left[{\bf U}(t,{\bf r})+\vibf(t,{\bf r})\right]+\viibf(t,{\bf r}).
\label{V_EXPANSION}
\eeq
Here, ${\bf V}={\bf U}+\vbf$ is the total fluid velocity, and the Kolmogorov turbulent 
velocities ${\bf U}$ are considered to be given and to be of the first order 
(Vainshtein 1970, Kulsrud \& Anderson 1992). 

Now, we substitute the above expansion formulas into 
equations~(\ref{v_EVOLUTION})--(\ref{b_EVOLUTION}). We find that 
the zero order equations are
\beq
\partial_t\,\Bo_\alpha = 0, 
\qquad
\partial_t\,\bo_\alpha = 0,
\qquad
\vo_\alpha = 0,
\qquad
\Vo_\alpha = 0,
\label{EVOLUTION_o}
\eeq
the first order equations are
\beq
\partial_t\,\Bi_\alpha &=& \Vi_{\alpha,\beta}\Bo_\beta 
-\Vi_\beta\Bo_{\alpha,\beta}\;,
\label{B_EVOLUTION_i}
\\
\partial_t\,\bi_\alpha &=& \Vi_{\alpha,\beta}\bo_\beta
-\Vi_{\beta,\gamma}\bbo_{\alpha\beta\gamma}-\Vi_\beta\bo_{\alpha,\beta}\;,
\label{b_EVOLUTION_i}
\\
\partial_t\,\vi_\alpha &=& -\:\Pi_{,\alpha}
+3\nu(\bbo_{\alpha\beta\mu\nu}\vi_{\mu,\nu})_{,\beta}
+3\nu(\bbo_{\alpha\beta\mu\nu}U_{\mu,\nu})_{,\beta}
-(1/5)\nu\triangle U_\alpha\;,
\;\,
\label{v_EVOLUTION_i}
\\
\vi_{\alpha,\alpha} &=& 0,
\label{v_DIV_i}
\\
\Vi_\alpha &=& U_\alpha+\vi_\alpha\;,
\label{Vi}
\eeq
and the second order equations are
\beq
\partial_t\,\Bii_\alpha \!\!&=&\!\! \Vi_{\alpha,\beta}\Bi_\beta 
+\Vii_{\alpha,\beta}\Bo_\beta -\Vi_\beta\Bi_{\alpha,\beta}
-\Vii_\beta\Bo_{\alpha,\beta}\;,
\label{B_EVOLUTION_ii}
\\
\partial_t\,\bii_\alpha \!\!&=&\!\! \Vi_{\alpha,\beta}\bi_\beta 
+\Vii_{\alpha,\beta}\bo_\beta -\Vi_{\beta,\gamma}\bbi_{\alpha\beta\gamma}
-\Vii_{\beta,\gamma}\bbo_{\alpha\beta\gamma} 
- \Vi_\beta\bi_{\alpha,\beta}
-\Vii_\beta\bo_{\alpha,\beta}\;,
\label{b_EVOLUTION_ii}
\\
\partial_t\,\vii_\alpha \!\!&=&\!\! -\,\Pii_{,\alpha}
+3\nu[\,\bbo_{\alpha\beta\mu\nu}\vii_{\mu,\nu}
+\bbi_{\alpha\beta\mu\nu}(\vi_{\mu,\nu}+U_{\mu,\nu})]_{,\beta}
-[\vi_\alpha U_\beta+U_\alpha\vi_\beta+\vi_\alpha\vi_\beta]_{,\beta},
\qquad
\label{v_EVOLUTION_ii}
\\
\vii_{\alpha,\alpha} \!\!&=&\!\! 0,
\label{v_DIV_ii}
\\
\Vii_\alpha \!\!&=&\!\! \vii_\alpha\;.
\label{Vii}
\eeq
Here, of course, the pressure is also expanded, $P=\Pi+\Pii$.

First, let us solve the first order equations~(\ref{v_EVOLUTION_i}) and~(\ref{v_DIV_i}).
We Fourier transform these equations in space and in time, ${\bf r}\rightarrow{\bf k}$
and $t\rightarrow\omega$, by making use of the discrete and the continuous Fourier 
transformations respectively [see eq.~(\ref{FOURIER})]. We have
\beq
(-i\omega+\Omega_{\rm rd})\viT_{{\bf k}\alpha}(\omega) \!\!&=&\!\! -ik_\alpha  \PiT_{\bf k}
+3\nu ik_\beta \!\!\sum_{{\bf k'}\atop{{\bf k}''=\,{\bf k}-{\bf k}'}}\!\!
ik'_\nu [\viT_{{\bf k}'\mu}+{\tilde U}_{{\bf k}'\mu}]\, 
\bboT_{{\bf k}''\alpha\beta\mu\nu} +(1/5)\nu k^2  {\tilde U}_{{\bf k}\alpha},
\qquad
\label{viT_EVOLUTION_PRESSURE}
\\
k_\alpha\viT_{{\bf k}\alpha}(\omega) \!\!&=&\!\! 0,
\label{viT_DIV}
\eeq
where $\viTbf_{\bf k}(\omega)$, $\PiT_{\bf k}(\omega)$, ${\bf\tilde U}_{\bf k}(\omega)$
and $\bboT_{{\bf k}\alpha\beta\mu\nu}$ are the Fourier coefficients. In the 
left-hand-side of equation~(\ref{viT_EVOLUTION_PRESSURE}) we add a damping term, 
$\Omega_{\rm rd}\viTbf_{\bf k}$, in order to account for the rotation of velocity 
vectors relative to the magnetic field unit vectors, see the discussion on 
page~\pageref{RD_EXPLANATION}. We estimate {\it the effective rotational damping rate} 
$\Omega_{\rm rd}$ on page~\pageref{OMEGA_DAMP}, and show it is a constant. In general, 
it may be a function of $k=|{\bf k}|$ and of 
$({\bf\b}\cdot{\bf\k})^2$.~\footnote{
It depends on the square of ${\bf\b}\cdot{\bf\k}$ because the Braginskii viscosity 
is invariant under reflection ${\bf\b}\rightarrow{}-{\bf\b}$ (gyrating ions 
``do not care'' about the exact direction of ${\bf\b}$).
\label{BRAGINSKII_IS_EVEN_IN_b}
}

Now, we multiply equation~(\ref{viT_EVOLUTION_PRESSURE}) on the left by tensor
$\delta^\perp_{\gamma\alpha}=\delta_{\gamma\alpha}-\k_\gamma\k_\alpha$ to 
eliminate the pressure term by making use of the incompressibility 
condition~(\ref{viT_DIV}). Interchanging indices, and using the symmetry of tensor 
$\bboT_{{\bf k}''\alpha\beta\mu\nu}$ with respect to its spatial indices, we obtain 
\beq
\MoT_{{\bf k}\alpha,{\bf k'}\beta}\viT_{{\bf k'}\beta} &=& \FiT_{{\bf k}\alpha}\;,
\label{viT_EVOLUTION}
\\
\MoT_{{\bf k}\alpha,{\bf k'}\beta} &=& (-i\omega+\Omega_{\rm rd})
\delta_{{\bf k},{\bf k'}}\delta_{\alpha\beta} 
+ 3\nu\delta^\perp_{\alpha\gamma}k_\mu k'_\nu \bboT_{{\bf k''}\gamma\mu\nu\beta}\;,
\quad{\bf k''}={\bf k}-{\bf k'},\qquad
\label{MoT_DEFINITION}
\\
\FiT_{{\bf k}\alpha} &=& \left[-\MoT_{{\bf k}\alpha,{\bf k'}\beta}
+\left(-i\omega+\Omega_{\rm rd}+\nu k^2/5\right)
\delta_{{\bf k},{\bf k'}}\delta_{\alpha\beta}\right]{\tilde U}_{{\bf k'}\beta}\;,
\label{FiT_DEFINITION}
\eeq
where we use convenient matrix notation, so that summation is implicitly assumed 
over repeated spatial indices and wave numbers. The matrix operator 
$\MoT_{{\bf k}\alpha,{\bf k'}\beta}(\omega)$ is of the zero order, while 
``the driving force'' $\FiT_{{\bf k}\alpha}(\omega)$ is of the first order.
If there exist inverse matrix $\MoT_{{\bf k}\alpha,{\bf k'}\beta}^{-1}$, then, 
using equations~(\ref{Vi}),~(\ref{viT_EVOLUTION}) and~(\ref{FiT_DEFINITION}), 
we obtain the Fourier coefficient of the first order total velocity $\Vibf$,
\beq 
\ViT_{{\bf k}\alpha}(\omega) &=& {\tilde U}_{{\bf k}\alpha} + \viT_{{\bf k}\alpha} 
= \left(-i\omega+\Omega_{\rm rd}+\nu k^2/5\right)
\MoT_{{\bf k}\alpha,{\bf k'}\beta}^{-1}\,{\tilde U}_{{\bf k'}\beta}(\omega)\;.
\label{ViT_SOLUTION}
\eeq
\clearpage

\begin{figure}[!t]
\vspace{7.0truecm}
\includegraphics{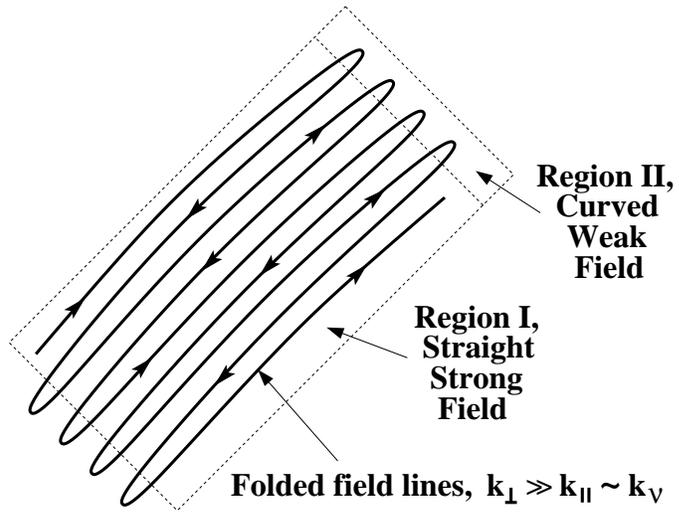}
\caption{
The folding structure of magnetic fields (for simplicity shown in two dimensions). 
The field is nearly straight and strong in Region I. The field is very curved but 
weak in Region II. 
}
\label{FIGURE_FOLDING}
\end{figure}
\clearpage

Now, let us consider an important case for which we can invert matrix 
$\MoT_{{\bf k}\alpha,{\bf k'}\beta}$, given by equation~(\ref{MoT_DEFINITION}).
Let $(\Bobf\cdot\nabla)\Bobf=0$, i.~e.~let at zero time the magnetic field  
vary only in the direction perpendicular to itself. This is equivalent to 
the magnetic field lines being initially straight (no curvature), and 
$\bbo_{\alpha\beta}=\bo_\alpha\bo_\beta\equiv{\rm const}$. This model of 
straight magnetic field lines is not as artificial as it seems at first glance, 
because of the following arguments. First, it is known that in the kinematic 
turbulent dynamo case the magnetic field lines have a folding pattern, shown 
in Figure~\ref{FIGURE_FOLDING}, (Schekochihin \etal 2002; Maron \& Cowley 2001). 
This folding pattern implies that in the bulk of the volume the field is strong 
and has small curvature, $k_\perp\gg k_\parallel\sim k_\nu$ ($k_\parallel$ and 
$k_\perp$ are the wave numbers parallel and perpendicular to the field lines), 
while in a small fraction of the volume the field is weak and curved, 
$k_\perp\sim k_\parallel\gg k_\nu$. The regions of weak and curved field, 
Region~II in Figure~\ref{FIGURE_FOLDING}, can be disregarded as long as we 
consider the volume averaged magnetic field energy and are not interested in the 
field curvature. As for the regions of strong magnetic field with small curvature, 
Region~I in Figure~\ref{FIGURE_FOLDING}, the field lines in these regions can be 
well approximated by the straight field lines on scales $k$ which satisfy 
$k\simgt k_\parallel\sim k_\nu$. Now, we make assumption that even for the 
magnetized turbulent dynamo case the magnetic field has a folding structure 
similar to that for the kinematic turbulent dynamo case. This assumption will be 
our {\it first working hypothesis}. It is based on the simulations of Maron and 
Cowley (2001), who found good indications for the magnetic field folding pattern 
in their numerical simulations of MHD turbulence with the Braginskii viscosity. 
Unfortunately, our calculational methods are not adequate to theoretically justify 
our first working hypothesis because of complications of the field curvature 
calculations. However, there exist a simple {\it reductio ad absurdum} theoretical 
argument, supporting the hypothesis, which is as follows. If the magnetic field 
were not folded, i.~e.~if $k_\perp\sim k_\parallel\simgt k_\nu$ everywhere in space, 
then the field would be isotropically tangled on viscous and subviscous scales. 
In this case the Braginskii viscosity could be averaged over the field direction, 
and it should quickly reduce to the isotropic effective viscosity 
$\nu_{\rm eff}=\nu/5$ [see eq.~(\ref{NU_EFF})]. As a result, in this case the 
magnetized dynamo should possess the properties of the kinematic dynamo 
with this effective viscosity, and would develop the folding structure of the 
magnetic field lines. 

Note that at a given fixed spatial point of the Eulerian system of coordinates 
the field curvature changes in time. Thus, even if the field is initially 
straight at this point, it may not remain straight in the future. In other words, 
as time goes on, sometimes the spatial point belongs to a region I and sometimes 
to a region II (this reflects the time intermittency and spatial convection
of the curvature). However, we are not interested in the magnetic energy 
evolution at each point of space, but instead, we are interested in the 
evolution of the total (spatially averaged) magnetic energy. Therefore, at any 
arbitrary moment of time we consider only those points of space that belong to 
all regions I at this time moment. These points give the dominant 
contribution to the total magnetic energy, and at these points the field lines 
can be considered to be straight. We apply our mathematical model of straight 
field lines to these points, and we find the change (the time derivative) 
of the total magnetic energy at the time moment we consider (see 
Section~\ref{MAGNETIC_SPECTRUM}). Because the moment of time is arbitrary, 
our final results give the correct evolution of the total magnetic energy in 
time.

Thus, from our first working hypothesis $\bbo_{\alpha\beta}\equiv{\rm const}$, 
we have $\bboT_{{\bf k''}\gamma\mu\nu\beta}=
\delta_{{\bf k''},0}\bo_\gamma\bo_\mu\bo_\nu\bo_\beta$, and can easily invert 
matrix~(\ref{MoT_DEFINITION}),
\beq
\MoT_{{\bf k}\alpha,{\bf k'}\beta}^{-1} &=& 
\frac{\delta_{{\bf k},{\bf k'}}}{-i\omega+\Omega_{\rm rd}} 
\left[\delta_{\alpha\beta}-
\frac{3\nu k^2\mu^2(1-\mu^2)}{-i\omega+\Omega_{\rm rd}+3\nu k^2\mu^2(1-\mu^2)}
\,\frac{\delta^\perp_{\alpha\gamma}\bo_\gamma\bo_\beta}{1-\mu^2}
\right].
\label{INVERSE_MATRIX}
\\
\mu &\define& {\bf\bobf}\cdot{\bf\k}.
\label{MU}
\eeq
We substitute this formula into equation~(\ref{ViT_SOLUTION}) and obtain
\beq
\ViT_{{\bf k}\alpha}(\omega) &=& \ViT'_{{\bf k}\alpha}(\omega)+\ViT''_{{\bf k}\alpha}(\omega)\;,
\label{ViT_F}
\\
\ViT'_{{\bf k}\alpha}(\omega) &=& 
\frac{-i\omega+\Omega_{\rm rd}+{\bar\Omega}}{-i\omega+\Omega_{\rm rd}} 
\left[\delta_{\alpha\beta}-\frac{\delta^\perp_{\alpha\gamma}\bo_\gamma\bo_\beta}{1-\mu^2}\right]
{\tilde U}_{{\bf k}\beta}(\omega)\;,
\label{ViT_F_FIRST}
\\
\ViT''_{{\bf k}\alpha}(\omega) &=& 
\frac{-i\omega+\Omega_{\rm rd}+{\bar\Omega}}{-i\omega+\Omega_{\rm rd}+2\Omega}\;
\frac{\delta^\perp_{\alpha\gamma}\bo_\gamma\bo_\beta}{1-\mu^2}\:
{\tilde U}_{{\bf k}\beta}(\omega)\;,
\label{ViT_F_SECOND}
\eeq
where, we introduce the following notations for the viscous damping frequencies
\beq
{\bar\Omega} \define (1/5)\nu k^2 = \nu_{\rm eff} k^2,
\qquad
\Omega \define (3/2)\nu k^2 \mu^2(1-\mu^2) = (15/2){\bar\Omega}\mu^2(1-\mu^2).
\label{OMEGA}
\eeq
The frequency $\Omega$ depends on $\mu^2$ because of the anisotropy of the Braginskii 
viscous stress tensor. The frequency ${\bar\Omega}$ is equal to $\Omega$ 
averaged over $\mu$, and represents the averaged rate of the Braginskii 
viscous dissipation [see eq.~(\ref{NU_EFF})].

Next, we calculate $\langle\ViT_{{\bf k}\alpha}(t)\rangle$ and 
$\langle\ViT_{{\bf k}\alpha}(t)\ViT_{{\bf k'}\beta}(t')\rangle$, which are the ensemble
averages of the ${\bf\Vi}$'s over all possible realizations of the turbulent motions. 
First, using formulas~(\ref{U_STATISTICS}),~(\ref{U_U_STATISTICS}) 
and~(\ref{ViT_F})--(\ref{ViT_F_SECOND}), we find
\beq
\langle\ViT_{{\bf k}\alpha}(\omega)\rangle \!\!&=&\!\! 0,
\label{ViT_F_AVERAGED}
\\
\langle\ViT_{{\bf k}\alpha}(\omega)\ViT_{{\bf k'}\beta}(\omega')\rangle \!\!&=&\!\!
\langle\ViT'_{{\bf k}\alpha}(\omega)\ViT'_{{\bf k'}\beta}(\omega')\rangle
+\langle\ViT''_{{\bf k}\alpha}(\omega)\ViT''_{{\bf k'}\beta}(\omega')\rangle
\nonumber\\
{} \!\!&=&\!\! J_{\omega k}\,\delta_{{\bf k'},-{\bf k}}\,\delta(\omega'+\omega)
\left[{\tilde H}_F(\omega;{\bar\Omega}+\Omega_{\rm rd},\Omega_{\rm rd})
\left(\delta^\perp_{\alpha\beta}-
\frac{\delta^\perp_{\alpha\gamma}\delta^\perp_{\beta\tau}\bo_\gamma\bo_\tau}
{1-\mu^2}\right)\right.
\nonumber\\
\!\!&&\!\!
{}+\left.{\tilde H}_F(\omega;{\bar\Omega}+\Omega_{\rm rd},2\Omega+\Omega_{\rm rd})\:
\frac{\delta^\perp_{\alpha\gamma}\delta^\perp_{\beta\tau}\bo_\gamma\bo_\tau}
{1-\mu^2}
\right],
\label{ViT_ViT_F_AVERAGED}
\\
{\tilde H}_F(\omega;q_1,q_2) \!\!&\define&\!\! 
\frac{\omega^2+q_1^2}{\omega^2+q_2^2}\,.
\label{H_F_T}
\eeq
Note that $\langle\ViT'_{{\bf k}\alpha}(\omega)\ViT''_{{\bf k'}\beta}(\omega')\rangle=0$, 
this implies that $\ViTbf'_{\bf k}$ and $\ViTbf''_{\bf k}$ are perpendicular on average. 
Second, we apply the inverse Fourier transformations in time, 
$\omega,\omega'\rightarrow t,t'$, to 
equations~(\ref{ViT_F_AVERAGED}),~(\ref{ViT_ViT_F_AVERAGED}). We have for the required
correlation functions of the total turbulent velocities
\beq
\langle\ViT_{{\bf k}\alpha}(t)\rangle \!\!&=&\!\! 0,
\label{ViT_AVERAGED_NEW}
\\
\langle\ViT_{{\bf k}\alpha}(t)\ViT_{{\bf k'}\beta}(t')\rangle \!\!&=&\!\!
\langle\ViT_{{\bf k}\alpha}(t)\ViT^*_{-{\bf k'}\beta}(t')\rangle =
\langle\ViT'_{{\bf k}\alpha}(t)\ViT'_{{\bf k'}\beta}(t')\rangle
+\langle\ViT''_{{\bf k}\alpha}(t)\ViT''_{{\bf k'}\beta}(t')\rangle
\nonumber\\
\!\!&=&\!\! \delta_{{\bf k'},-{\bf k}}\;
\left[H_F(t-t';{\bar\Omega}+\Omega_{\rm rd},\Omega_{\rm rd})
\left(\delta^\perp_{\alpha\beta}-
\frac{\delta^\perp_{\alpha\gamma}\delta^\perp_{\beta\tau}\bo_\gamma\bo_\tau}
{1-\mu^2}\right)\right.
\nonumber\\
\!\!&&\!\!\left.\quad{}
+H_F(t-t';{\bar\Omega}+\Omega_{\rm rd},2\Omega+\Omega_{\rm rd})\:
\frac{\delta^\perp_{\alpha\gamma}\delta^\perp_{\beta\tau}\bo_\gamma\bo_\tau}{1-\mu^2}
\right],
\label{ViT_ViT_AVERAGED_NEW}
\eeq
where the function
\beq
H_F(t-t';q_1,q_2) = \frac{1}{2\pi}\! \int_{-\infty}^\infty \!
J_{\omega k}{\tilde H}_F(\omega;q_1,q_2)\,e^{-i\omega(t-t')} d\omega
{}={} \frac{J_{0k}}{2\pi}\! \int_{-\infty}^\infty 
\frac{\omega^2+q_1^2}{\omega^2+q_2^2}\,\frac{\cos{[\omega(t-t')]}}{1+\tau^2\omega^2}\,d\omega
\quad
\label{H_F}
\eeq
is the inverse Fourier transform of function $J_{\omega k}{\tilde H}_F(\omega;q_1,q_2)$, 
and depends only on the absolute value of the time difference $t-t'$. In 
equation~(\ref{H_F}) we use formulas~(\ref{J_OMEGA_K}) and~(\ref{H_F_T}).

The main unknown quantity is the effective rotational damping coefficient 
$\Omega_{\rm rd}$, which we introduced in equation~(\ref{viT_EVOLUTION_PRESSURE}) 
as the $\Omega_{\rm rd}\viTbf_{\bf k}$ term\label{RD_EXPLANATION}. Let us consider 
it in more detail. We start with the calculation of the ensemble average of $\Vibf$ 
squared~\footnote{
Note that $\langle\Vibf\rangle=0$ because there is no preferred direction. 
Of course, there is a preferred axis in space, which is along the magnetic field 
unit vector, see also the footnote~\ref{BRAGINSKII_IS_EVEN_IN_b} on 
page~\pageref{BRAGINSKII_IS_EVEN_IN_b}.
}.
We have
\beq
\langle\Vibf^2\rangle &=&
\langle\Vi_\alpha(t,{\bf r})\Vi_\alpha(t,{\bf r})\rangle
= \sum\limits_{{\bf k},{\bf k'}}
\langle\ViT_{{\bf k}\alpha}(t)\ViT_{{\bf k'}\alpha}(t)\rangle\,
e^{i({\bf k}+{\bf k'}){\bf r}}
=\sum\limits_{{\bf k}}\langle|\ViTbf_{\bf k}|^2\rangle
\label{Vi_SQUARED}
\eeq
[see eq.~(\ref{ViT_ViT_AVERAGED_NEW})]. Using equations~(\ref{ViT_ViT_AVERAGED_NEW}) 
and~(\ref{H_F}), we obtain
\beq
\langle|\ViTbf_{\bf k}|^2\rangle &=&
H_F(0;{\bar\Omega}+\Omega_{\rm rd},\Omega_{\rm rd})
+H_F(0;{\bar\Omega}+\Omega_{\rm rd},2\Omega+\Omega_{\rm rd})
\nonumber\\
{}&=& \frac{J_{0k}}{\tau}
+\frac{J_{0k}}{2}\frac{({\bar\Omega}+\Omega_{\rm rd})^2-\Omega_{\rm rd}^2}
{\Omega_{\rm rd}(1+\tau \Omega_{\rm rd})}
+\frac{J_{0k}}{2}\frac{({\bar\Omega}+\Omega_{\rm rd})^2-(2\Omega+\Omega_{\rm rd})^2}
{(2\Omega+\Omega_{\rm rd})[1+\tau(2\Omega+\Omega_{\rm rd})]}.
\label{ViT_SQUARED}
\eeq
According to equations~(\ref{Vi_SQUARED}) and~(\ref{ViT_SQUARED}), we see that 
if we set the rotational damping rate $\Omega_{\rm rd}$ to zero, then 
the ensemble average of the first order velocity squared would become infinite, 
$\langle\Vibf^2\rangle\to\infty$ as $\Omega_{\rm rd}\to 0\,$! Let us try to 
understand this divergence problem, and see how we can avoid it.

First, following the derivation of formula~(\ref{ViT_SQUARED}) from 
equation~(\ref{ViT_ViT_AVERAGED_NEW}), it is easy to see that the divergence of 
$\langle\Vibf^2\rangle$ occurs only due to the divergence of the 
$\ViT'_{{\bf k}\alpha}$ modes. In other words, 
$\langle|\ViTbf'_{\bf k}|^2\rangle\to\infty$ as $\Omega_{\rm rd}\to 0$, while 
$\langle|\ViTbf''_{\bf k}|^2\rangle$ stays finite~\footnote{
Of course, in the degenerate cases, when ${\bf\k}\perp\bobf$ (i.~e.~$\mu^2=0$) 
or ${\bf\k}\parallel\bobf$ (i.~e.~$1-\mu^2=0$), both $\ViTbf'_{\bf k}$ 
and $\ViTbf''_{\bf k}$ modes become infinite as $\Omega_{\rm rd}\to 0$.
}.
Second, let us refer to equations~(\ref{ViT_F_FIRST}) and~(\ref{ViT_F_SECOND}). 
On one hand, we have $\k_\alpha\ViT'_{{\bf k}\alpha} = 0$ and 
$\k_\alpha\ViT''_{{\bf k}\alpha} = 0$, as it should be because plasma velocities 
are incompressible. On the other hand, we have $\bo_\alpha\ViT'_{{\bf k}\alpha} = 0$ 
and $\bo_\alpha\ViT''_{{\bf k}\alpha} \ne 0$. Thus, the divergent velocity modes, 
$\ViTbf'_{\bf k}$, are perpendicular to both vectors ${\bf\k}$ and ${\bf\bo}$. 
At the same time, the other, non-divergent modes, $\ViTbf''_{\bf k}$, have nonzero 
components along ${\bf\bo}$. Third, let us calculate the ensemble averaged Braginskii 
viscous dissipation into heat (Braginskii 1965). We have
\beq
\langle Q_{\rm vis}\rangle &=& \langle\pi_{\alpha\beta}V_{\alpha,\beta}\rangle
={}-3\nu\langle\Vi_{\alpha,\beta}\Vi_{\gamma,\tau}\rangle\bbo_{\alpha\beta\gamma\tau}
=3\nu\sum\limits_{{\bf k},{\bf k'}} k_\beta k'_\tau
\langle\ViT_{{\bf k}\alpha}\ViT_{{\bf k'}\gamma}\rangle\,
\bbo_{\alpha\beta\gamma\tau}\, e^{i({\bf k}+{\bf k'}){\bf r}}
\nonumber\\
{}&=& {}-3\nu\sum\limits_{\bf k} k^2\mu^2\:
\bo_\alpha\langle\ViT''_{{\bf k}\alpha}\ViT''_{-{\bf k}\gamma}\rangle\bo_\gamma
= {}-2\sum\limits_{\bf k} 
\Omega\,H_F(0;{\bar\Omega}+\Omega_{\rm rd},2\Omega+\Omega_{\rm rd}),
\eeq
where we keep only the first order velocities, and make use of formulas
$\bbo_{\alpha\beta\gamma\tau}={\rm const}$, $\bo_\alpha\ViT'_{{\bf k}\alpha} = 0$, 
and of equations~(\ref{PI}),~(\ref{MU}),~(\ref{OMEGA}),~(\ref{ViT_ViT_AVERAGED_NEW}). 
We see that only the $\ViTbf''_{\bf k}$ velocity modes are dissipated by the 
Braginskii viscous forces, and the dissipation is proportional to 
$2\Omega=3\nu k^2\mu^2(1-\mu^2)$. 

As a result, we conclude that the divergence of the first order velocities 
happens because the $\ViTbf'_{\bf k}$ velocity modes, which are perpendicular 
to ${\bf\bo}$, are not damped by the Braginskii viscous dissipation. On the 
other hand, it is clear that the $\ViTbf'_{\bf k}$ modes can not be infinite. 
What are the physical mechanisms which limit them? To answer this question, 
let us note that velocity modes are non-linearly coupled with each other via 
the inertial term, $({\bf V\cdot\nabla}){\bf V}$, of the MHD 
equation~(\ref{EQUATION_FOR_V}). This non-linear coupling transforms the 
divergent $\ViTbf'_{\bf k}$ velocity modes into other modes, which are then 
viscously dissipated. This transformation can be viewed as a continuous 
rotation of velocity vectors relative to the magnetic field unit vector 
${\bf\b}$. Indeed, at a given point of space we can go to a reference frame 
that rotates together with ${\bf\b}$. In this rotating frame there 
exist Coriolis forces, which act on velocity vectors and force them to rotate 
relative to the non-rotating field vector. These Coriolis forces are caused 
by the non-linear coupling of velocity modes via the inertial term. As a 
result, a divergent velocity mode, perpendicular to ${\bf\b}$ and not viscously 
damped, eventually rotates out of its initial direction and is transformed into 
a damped mode. We call this process the ``effective rotational damping''.
Of course, it operates only on the viscous scales, on which the Braginskii 
viscous dissipation is significant. On larger scales the viscous dissipation 
is small and the rotation of velocities does not make any difference. 

We could regard $\Omega_{\rm rd}$ as a parameter to be determined by numerical
simulations. Nevertheless, it is of interest to attempt to estimate it from
a physical argument. First, the square of the angular velocity of the rotation 
can be estimated as $\omega_{\rm rot}^2\sim(1/3)\tau_\nu^{-2}$, where $\tau_\nu$ 
is the velocity decorrelation time on the viscous scale. The factor $1/3$ in 
this equation comes from the fact that only one of the three angular velocity 
components contributes to the deviation of a divergent velocity mode from its 
original direction perpendicular to the field unit vector 
${\bf\b}$~\footnote{
This is the component along the vector product ${\bf\b}\times{\bf V}$. 
}.
Second, the typical Braginskii viscous damping rate is clearly about 
$1/\tau_\nu\sim\nu_{\rm eff}k_\nu^2$, and is larger than the angular velocity 
of the rotation. Let assume it is much larger, 
$\omega_{\rm rot}^2\tau_\nu^2\sim 1/3\ll 1$. Then we can suppose that after 
the divergent velocity mode, in a time interval $\Delta t$, rotates by an 
angle $\Delta\phi$ relative to its original direction perpendicular to 
${\bf\b}$, only the projection of the velocity mode on the plane perpendicular 
to ${\bf\b}$ survives, and all other velocity components are immediately 
viscously dissipated. As a result, the effective rotational damping rate, 
$\Omega_{\rm rd}$, can be estimated as follows.
\beq
\frac{dV}{dt} &\sim& \frac{\Delta V}{\Delta t}
\sim \frac{V(\cos{\Delta\phi}-1)}{\Delta t} 
\sim {}-\frac{V}{2\Delta t}\:{\Delta\phi}^2
\sim {}-\frac{V}{2\Delta t}\;(\omega_{\rm rot}\tau_\nu)^2\frac{\Delta t}{\tau_\nu}
\nonumber\\
&\sim& {}-(1/6)\tau_\nu^{-1}V \sim (1/6)\nu_{\rm eff} k_\nu^2 V
\sim (1/30)\nu k_\nu^2 V = {}-\Omega_{\rm rd}V,
\label{ROTATIONAL_DAMPING}
\\
\Omega_{\rm rd} &=& (1/6)\nu_{\rm eff} k_\nu^2 = (1/30)\nu k_\nu^2.
\label{OMEGA_DAMP}
\eeq
To obtain the last result in the first line of equation~(\ref{ROTATIONAL_DAMPING}), 
we use the random-walk approximation for the estimate of 
${\Delta\phi}^2$.~\footnote{
We again assume that $\omega_{\rm rot}^2\tau_\nu^2\sim 1/3\ll 1$. Next we 
choose such time interval $\Delta t\gg\tau_\nu$, that 
$\Delta\phi\sim\omega_{\rm rot}\Delta t\ll 1$. Then we can simultaneously expand 
the $\cos{\Delta\phi}$ in eq.~(\ref{ROTATIONAL_DAMPING}) and use the random-walk 
approximation to estimate ${\Delta\phi}^2$.
}

Let us summarize our discussion of the effective rotational damping. As we said, 
this physical damping is associated with the non-linear coupling of velocity modes, 
which leads to the rotation of velocities relative to the magnetic field vectors. 
The effective rotational damping is very important because it limits the velocity 
modes which are perpendicular to the magnetic field vectors, and therefore, are 
undamped by the Braginskii viscous forces. The non-linear coupling of the velocity 
modes is hard to deal with directly. In particular, the MHD non-linear inertial 
terms do not appear in our first-order equation~(\ref{viT_EVOLUTION_PRESSURE}). 
As a result, to avoid the divergence problem for the undamped velocity modes, we 
incorporate the non-linear mode coupling and the associated rotational damping 
into our equations in a simple way, as the $\Omega_{\rm rd}\viTbf_{\bf k}$ damping 
term in the left-hand-side of equation~(\ref{viT_EVOLUTION_PRESSURE}). This is our 
{\it second working hypothesis}. Note, that the $\Omega_{\rm rd}\viTbf_{\bf k}$ 
term is isotropic, and therefore, it damps not only the divergent velocity 
modes, perpendicular to the field vector, but all velocity modes. This is not a 
serious problem though, because the rotational damping is smaller than the 
Braginskii damping by a factor ${}\sim 1/6$ [see eq.~(\ref{OMEGA_DAMP})], and our 
results should be valid within a factor of order unity. Also note that on 
scales larger than the viscous scale the rotational damping does not operate. 
However, on those large scales the turbulence is Kolmogorov, ${\bf V}={\bf U}$, 
and the back reaction velocities $\vbf$ are zero 
anyway~\footnote{ 
In particular, note that the driving force~(\ref{FiT_DEFINITION}), which is 
${}\propto k^2$, becomes small on large scales.
}.

So far we have considered only the first order velocity, $\Vibf$, and have found its 
statistics, given by equations~(\ref{ViT_AVERAGED_NEW})--(\ref{H_F}) 
and~(\ref{OMEGA_DAMP}). In order to find the second order velocity, $\Viibf$, we need 
to solve the complicated second order equations~(\ref{v_EVOLUTION_ii})--(\ref{Vii}). 
Fortunately, we will need only the ensemble average of the second order velocity, 
$\langle\ViiTbf_{\bf k}\rangle$. It turns out that in our case of a straight 
initial field, $\bbo_{\alpha\beta}={\rm const}$, which we consider here, this 
average is zero,
\beq
\langle\ViiT_{{\bf k}\alpha}(t)\rangle &=& 0,
\label{Vii_AVERAGE_NEW}
\eeq
(Malyshkin 2001). The reason for this simple result is that $\bbo_{\alpha\beta}$ 
is constant in space, and the Kolmogorov turbulence, ${\bf U}$, is statistically
homogeneous. Therefore, the ensemble averages of the terms in the brackets $[...]$ 
in equation~(\ref{v_EVOLUTION_ii}) are constant in space, and their spatial 
derivatives are zero. As a result, the ensemble averaged velocity 
$\langle\ViiT_{{\bf k}\alpha}\rangle=\langle\viiT_{{\bf k}\alpha}\rangle$ 
is also zero.

To conclude this section, let us integrate equation~(\ref{H_F}) in time,
and obtain the formulas
\beq
\int_0^t\! dt'\!\!\int_0^{t'}\! H_F(t'-t'')dt'' \!\!&=&\!\!
\frac{J_{0k}}{\pi}\!\int_{-\infty}^\infty 
\frac{\omega^2+q_1^2}{\omega^2+q_2^2}\,
\frac{\sin^2{(\omega t/2)}}{1+\tau^2\omega^2}\,\frac{d\omega}{\omega^2}
= \frac{J_{0k}}{2}\left[t-\tau(1-e^{-t/\tau})\right]
\nonumber\\
{}\!\!&+&\!\!
\frac{J_{0k}}{2}\frac{q_1^2-q_2^2}{1-\tau^2q_2^2}
\left[\frac{q_2t-1+e^{-q_2t}}{q_2^3}-\tau^2t+\tau^3(1-e^{-t/\tau})\right]
\rightarrow \frac{J_{0k}}{2}\frac{q_1^2}{q_2^2}\,t\,,\qquad
\label{H_F_INTEGRAL_T'T}
\\
\int_0^t\! dt'\!\!\int_0^{t}\! H_F(t'-t'')dt'' \!\!&=&\!\!
2\int_0^t\! dt'\!\!\int_0^{t'}\! H_F(t'-t'')dt''
\rightarrow J_{0k}\,\frac{q_1^2}{q_2^2}\,t\,,
\label{H_F_INTEGRAL_TT}
\eeq
which we will use below. Here, the integrals over $\omega$ can be done by closing 
the integration contours in the complex plane and by evaluating the residues. 
The final answers in these formulas, written after the right arrows 
``$\rightarrow$'', give the results in the limit $t\gg \tau,\,q_2^{-1}$.

\section
{Energy Spectrum of Random Magnetic Fields}
\label{MAGNETIC_SPECTRUM}

In this section we use equations~(\ref{ViT_AVERAGED_NEW}),~(\ref{ViT_ViT_AVERAGED_NEW})
and~(\ref{Vii_AVERAGE_NEW}), which give the statistics of the turbulent velocities, to 
derive the evolution of the magnetic energy in the magnetized turbulent dynamo theory.

\subsection
{The Growth of the Total Magnetic Energy}
\label{MAGNETIC_ENERGY}

The volume averaged and ensemble averaged magnetic energy per unit mass is
\beq
{\cal E} &\define& \frac{1}{L^3}\int\frac{\langle B^2\rangle}{8\pi\rho}\,d^3{\bf r}
=\frac{1}{8\pi\rho}\Big\langle\BST_{{\bf k}=0}\Big\rangle\:,
\label{AVERAGED_MAGNETIC_ENERGY}
\eeq
where $\rho$ is the plasma density, and $\BST_{{\bf k}=0}$ is the ${\bf k}=0$ 
Fourier coefficient of the magnetic field strength squared, $B^2$. To find 
${\cal E}(t)$, it is convenient to introduce a symmetric tensor
$B_{\alpha\beta}\define B_\alpha B_\beta=B^2\,b_{\alpha\beta}$.
The differential equation for $B_{\alpha\beta}$ follows from 
equation~(\ref{B_EVOLUTION}),
\beq
\partial_t B_{\alpha\beta} = B_\alpha\partial_t B_\beta + B_\beta\partial_t B_\alpha
=V_{\alpha,\gamma} B_{\beta\gamma}+V_{\beta,\gamma} B_{\alpha\gamma}
-V_\gamma B_{\alpha\beta,\gamma}\;.
\label{BB_EVOLUTION}
\eeq
Now, we simultaneously solve this equation and equation~(\ref{B_STRENGTH_EVOLUTION}) 
by making use of the quasi-linear expansion procedure, described in 
Section~\ref{VELOCITIES}. First, we write
\beq
B^2(t) = \BSo+\BSi(t)+\BSii(t),
\qquad
B_{\alpha\beta}(t) = \Bo_{\alpha\beta}+\Bi_{\alpha\beta}(t).
\label{BS_EXPANSION}
\eeq
Second, we substitute these expansion formulas into 
equations~(\ref{B_STRENGTH_EVOLUTION}) and~(\ref{BB_EVOLUTION}). 
We find that the first order equations are
\beq
\partial_t\BSi &=& 
2\,\Vi_{\alpha,\beta} \Bo_{\alpha\beta} - \Vi_\beta \BSo_{,\beta}\,,
\label{BS_FIRST_ORDER_EVOLUTION}
\\
\partial_t \Bi_{\alpha\beta} &=&
\Vi_{\alpha,\gamma}\Bo_{\beta\gamma}+\Vi_{\beta,\gamma}\Bo_{\alpha\gamma}
-(\Vi_\gamma\Bo_{\alpha\beta})_{,\gamma}\,,
\label{BB_FIRST_ORDER_EVOLUTION}
\eeq
and the second order equation for $\BSii(t)$ is
\beq
\partial_t\BSii &=& 
2\,\Vi_{\alpha,\beta} \Bi_{\alpha\beta}
+2\,\Vii_{\alpha,\beta} \Bo_{\alpha\beta} 
-(\Vi_\beta \BSi)_{,\beta}
-(\Vii_\beta \BSo)_{,\beta}\,.
\label{BS_SECOND_ORDER_EVOLUTION}
\eeq
Third, we integrate equation~(\ref{BS_FIRST_ORDER_EVOLUTION}) 
in time (with the zero initial conditions) and ensemble average the result. Using 
equation~(\ref{ViT_AVERAGED_NEW}), we obviously obtain $\langle\BSi(t)\rangle = 0$.
Fourth, we integrate equation~(\ref{BB_FIRST_ORDER_EVOLUTION}) in time, and then 
Fourier transform the result in space, ${\bf r}\rightarrow{\bf k}$. We have
\beq
\BiT_{{\bf k}\alpha\beta}(t) = i\Big[k'_\gamma(\delta_{\alpha\tau}\bbo_{\beta\gamma}
+\delta_{\beta\tau}\bbo_{\alpha\gamma})-k_\tau\bbo_{\alpha\beta}\Big]
\int_0^t \!\sum_{{\bf k'}\atop{{\bf k}''=\,{\bf k}-{\bf k}'}}\!\!\!
\ViT_{{\bf k'}\tau}(t')\,\BSoT_{\bf k''}\,dt'.
\label{BBiT}
\eeq
Here, we use $B_{\alpha\beta}=B^2\,b_{\alpha\beta}$ and $\bbo_{\alpha\beta}={\rm const}$.
(This last formula is our first working hypothesis). 
Fifth, we integrate the second order equation~(\ref{BS_SECOND_ORDER_EVOLUTION}) 
in time. Then we ensemble average the result and Fourier transform it in space,
setting ${\bf k}$ to zero. Using 
equations~(\ref{ViT_ViT_AVERAGED_NEW}),~(\ref{Vii_AVERAGE_NEW}) and~(\ref{BBiT}), 
we obtain
\beq
\left\langle\BSiiT_{{\bf k}=0}(t)\right\rangle \!\!&=&\!\!
2i\int_0^t \sum_{\bf k}
k_\beta\,\Big\langle\ViT_{{\bf k}\alpha}(t')\,\BiT_{-{\bf k}\alpha\beta}(t')\Big\rangle\,dt'
\nonumber\\
\!\!&=&\!\! 2\,\BSoT_{{\bf k}=0} \int_0^t dt'\int_0^{t'} \sum_{\bf k}
\mu^2k^2\, \langle\ViT_{{\bf k}\alpha}(t')\ViT_{-{\bf k}\alpha}(t'')\rangle
\,dt'' \:=\: 2\,\BSoT_{{\bf k}=0}\,\sum_{\bf k}\mu^2k^2\, 
\nonumber\\
\!\!&&\!\!{}\times
\int_0^t dt'\int_0^{t'} \Big[H_F(t'-t'';{\bar\Omega}+\Omega_{\rm rd},\Omega_{\rm rd})
+H_F(t'-t'';{\bar\Omega}+\Omega_{\rm rd},2\Omega+\Omega_{\rm rd})\Big]
\,dt''
\nonumber\\
{}&=& 2\gamma t\,\BSoT_{{\bf k}=0}\:,
\label{BSiiT_AVERAGED}
\eeq
where 
\beq
&& \gamma = \pi{\left(\frac{L}{2\pi}\right)}^{\!3} \int_0^\infty k^4 J_{0k}\,dk
\int_{-1}^1 \mu^2\left(1+{\bar\Omega}/\Omega_{\rm rd}\right)^2\,
\left[\,1+\left(1+2\Omega/\Omega_{\rm rd}\right)^{-2}\,\right]d\mu\,,
\qquad
\label{GAMMA}
\\
&& {\bar\Omega}/\Omega_{\rm rd} = 6k^2/k_\nu^2\,,
\qquad
2\Omega/\Omega_{\rm rd} = 90\,(k^2/k_\nu^2)\,\mu^2(1-\mu^2)\,.
\label{OMEGA_RATIO}
\eeq
Here, we also use equation~(\ref{H_F_INTEGRAL_T'T}) in the limit $t\gg\tau$, replace 
the summation over ${\bf k}$ by integration, making use of 
$d^3{\bf k}=2\pi k^2\,dk\,d\mu$, and use equations~(\ref{OMEGA}),~(\ref{OMEGA_DAMP}).

Now, we derive the differential equation for the averaged magnetic energy $\cal E$. 
Following Kulsrud and Anderson (1992), we choose $t$ small enough for the 
quasi-linear expansion to be valid, but large enough for the limit $t\gg\tau$ 
to be satisfied. This is very similar to an assumption that the turbulent 
velocities are $\delta$-correlated in time (Kazantsev 1968), the assumption 
generally used in the dynamo theories (see the discussion in the end of this section).
As a result, using equations~(\ref{BS_EXPANSION}) and~(\ref{BSiiT_AVERAGED}), 
we obtain
\beq
\partial_t\left\langle\BST_{{\bf k}=0}\right\rangle =
\frac{1}{t}\left\langle\BST_{{\bf k}=0}(t)-\BST_{{\bf k}=0}(0)\right\rangle
=\frac{1}{t}\left\langle\BSiT_{{\bf k}=0}+\BSiiT_{{\bf k}=0}\right\rangle
=\frac{1}{t}\left\langle\BSiiT_{{\bf k}=0}\right\rangle
=2\gamma\BSoT_{{\bf k}=0}\,,
\label{ENERGY_FINITE_DIFFERENCE}
\eeq
and using equation~(\ref{AVERAGED_MAGNETIC_ENERGY}), we finally obtain
\beq
\partial{\cal E}/\partial t = 2\gamma{\cal E}.
\label{ENERGY_GROWTH}
\eeq

According to this last equation, the magnetic energy grows exponentially in 
time, the same way as it does in the kinematic dynamo theory (Kulsrud \& Anderson 
1992). However, the growth rate $\gamma$, given by equation~(\ref{GAMMA}) in the 
magnetized turbulent dynamo case, is different from the growth rate $\gamma_{\rm o}$ 
in the kinematic dynamo case. The latter can easily by obtained by taking the limit 
$\Omega_{\rm rd}\rightarrow\infty$ in equation~(\ref{GAMMA}) and by changing the 
viscous cutoff wave number in equation~(\ref{J_ZERO_K}) to the standard one, 
$k_\nu=(U_0/k_0\nu)^{3/4}k_0$. In this limit, the back-reaction velocities are zero 
(because they are totally damped), the turbulence is Kolmogorov, and 
equation~(\ref{GAMMA}) reduces to the corresponding formula of Kulsrud and 
Anderson (1992), as one might expect. 

The integrals in equation~(\ref{GAMMA}) can be carried out numerically 
(Malyshkin 2001). The result is
\beq
&& \gamma \approx 8.5\,\left(U_0L/\nu\right)^{1/2}(U_0/L),
\qquad
\gamma^{-1} \approx 10^6\:{\rm yrs}\, \left(\xi/10\right)^{1/2}
\left(L/0.2\,{\rm Mpc}\right)^{3/2},
\label{GAMMA_RESULT}
\\
&& \gamma\,t_{\rm collapse} \approx \gamma(L/U_0)\approx
10^3\, \left(\xi/10\right)^{-1/2}
\left(M/10^{12}\,{\rm M}_\odot\right)^{-1/2},
\label{GAMMA_COLLAPSE_TIME}
\\
&&\gamma/\gamma_{\rm o} \approx 10.
\label{GAMMA_RATIO}
\eeq
Here, $t_{\rm collapse}$ is the protogalaxy collapse time, $\xi$ is the ratio 
of the total mass $M$ to the baryon mass, $L$ is the protogalaxy size, and the 
temperature is assumed to be virial. Equation~(\ref{GAMMA_RATIO}) predicts that 
the magnetic energy growth rate in the magnetized dynamo theory is up to ten 
time larger than that in the kinematic dynamo theory. Two different effects 
contribute to this difference. First, the effective viscosity in the magnetized 
dynamo case is smaller than the molecular viscosity, $\nu_{\rm eff}=\nu/5$. This 
effect makes the growth rate larger by a factor of $\sqrt{5}$ (this factor was 
included in Kulsrud \etal 1997). The rest of the contribution comes from the 
local anisotropy of the turbulent velocities in a strongly magnetized plasma. 

Note that according to equations~(\ref{TAU}) and~(\ref{GAMMA_RESULT}), the 
magnetic field growth time, $\gamma^{-1}=0.12(U_0L/\nu)^{-1/2}(L/U_0)$, is 
approximately $40\%$ smaller than the eddy turnover time on the viscous scale, 
$\tau(k_\nu)\sim 0.18(U_0L/\nu)^{-1/2}(L/U_0)$.~\footnote{
In the kinematic dynamo theory $\tau(k_\nu)\sim(1/3)\gamma^{-1}$ (Kulsrud \& 
Anderson 1992).
}
Thus, the quasi-linear expansion of the MHD equations in time may not be fully
compatible with the assumption of the $\delta$-time correlation of the turbulent 
velocities. On the other hand, the effects of the finite velocity correlation time 
should decrease the magnetic energy growth rate by a factor of order two (this 
reduction was found in the kinematic turbulent dynamo theory by Schekochihin \& 
Kulsrud, 2001). As a result of this reduction, the expansion should become better 
justified. Thus, including the finite time correlation effects into our theory 
would be important but not vital, since our calculation predicts a very large 
energy growth rate anyway [see eqs.~(\ref{GAMMA_RESULT})--(\ref{GAMMA_RATIO})].

\subsection
{The Mode Coupling Equation for the Magnetic Energy Spectrum}
\label{MODE_COUPLING_EQUATION}

The ensemble averaged magnetic energy spectrum is
\beq
M(t, k) = \frac{1}{4\pi\rho}{\left(\frac{L}{2\pi}\right)}^{\!3}
\int k^2\,\langle|{\bf\tilde B}(t,{\bf k})|^2\rangle\,d^2{\bf\k}, 
\label{M_K}
\eeq
where the integration is carried out over all directions of ${\bf\k}={\bf k}/k$, and 
${\bf\tilde B}_{\bf k}$ is the Fourier coefficient of the magnetic field, ${\bf B}$.
The total magnetic energy, given by equation~(\ref{AVERAGED_MAGNETIC_ENERGY}), 
is obviously 
\beq
{\cal E}=\frac{1}{2}\int_0^\infty M(t,k)\,dk.
\label{TOTAL_MAGNETIC_ENERGY}
\eeq
To find the evolution of $M(t, k)$, we use the quasi-linear expansion 
formula~(\ref{B_EXPANSION}) for ${\bf B}$. We write the ensemble averaged 
square of the magnetic field Fourier coefficient, up to the second order,
\beq
\langle|{\bf\tilde B}_{\bf k}(t)|^2\rangle &=& |\BoTbf_{\bf k}(t)|^2 
+[\langle\BiT_{{\bf k}\alpha}(t)\rangle\BoT^*_{{\bf k}\alpha}(t)+{\rm c.c.}]
+\langle|\BiTbf_{\bf k}(t)|^2\rangle
+[\langle\BiiT_{{\bf k}\alpha}(t)\rangle\BoT^*_{{\bf k}\alpha}(t)+{\rm c.c.}]
\nonumber\\
&=& |\BoTbf_{\bf k}(t)|^2 +\langle|\BiTbf_{\bf k}(t)|^2\rangle
+[\langle\BiiT_{{\bf k}\alpha}(t)\rangle\BoT^*_{{\bf k}\alpha}(t)+{\rm c.c.}].
\label{BkS_EXPANSION}
\eeq
Here, ${\rm c.c.}$ is the complex conjugate, and we use the fact that the ensemble 
averaged first order magnetic field is zero, $\langle\BiTbf_{\bf k}\rangle = 0$,
[this follows from eqs.~(\ref{B_EVOLUTION_i}) and~(\ref{ViT_AVERAGED_NEW})]. 
We calculate the two last terms in the right-hand-side of 
equation~(\ref{BkS_EXPANSION}) separately.

We start with calculation of the $\langle|\BiTbf_{\bf k}(t)|^2\rangle$ term. 
First, we integrate equation~(\ref{B_EVOLUTION_i}) in time, and then Fourier 
transform the result in space, ${\bf r}\rightarrow{\bf k}$. We have
\beq
\BiT_{{\bf k}\chi}(t) = ik_\gamma(\delta_{\chi\alpha}\delta_{\gamma\delta}
-\delta_{\gamma\alpha}\delta_{\chi\delta})\int_0^t 
\sum_{{\bf k'}\atop{{\bf k''}=\,{\bf k}-{\bf k'}}}
\ViT_{{\bf k''}\alpha}(t')\,\BoT_{{\bf k'}\delta}\,dt',
\label{BiT}
\eeq
where we use the divergence free conditions $k'_\alpha\BoT_{{\bf k'}\alpha}=0$ 
and $k''_\alpha\ViT_{{\bf k''}\alpha}=0$. Using this equation and its 
complex conjugate, we obtain
\beq
\langle|\BiTbf_{\bf k}(t)|^2\rangle &=& k_\gamma k_\tau 
(\delta_{\chi\alpha}\delta_{\gamma\delta}-\delta_{\gamma\alpha}\delta_{\chi\delta})
(\delta_{\chi\beta}\delta_{\tau\eta}-\delta_{\tau\beta}\delta_{\chi\eta})
\nonumber\\
&&{}\times \int_0^t\!\int_0^t
\!\!\sum_{{\bf k'}\atop{{\bf k''}=\,{\bf k}-{\bf k'}}}
\sum_{{\bf k'''}\atop{{\bf k^{iv}}=\,{\bf k}-{\bf k}'''}}\!\!\!\!
\BoT_{{\bf k'}\delta}\BoT^*_{{\bf k'''}\eta}
\langle\ViT_{{\bf k''}\alpha}(t')\ViT^*_{{\bf k^{iv}}\beta}(t'')\rangle
\,dt'dt''
\nonumber\\
&=&
k^2\,\k_\gamma\k_\tau (\delta_{\alpha\beta}\delta_{\gamma\delta}\delta_{\tau\eta}
-\delta_{\alpha\eta}\delta_{\beta\tau}\delta_{\gamma\delta}
-\delta_{\alpha\gamma}\delta_{\beta\delta}\delta_{\tau\eta}
+\delta_{\alpha\gamma}\delta_{\beta\tau}\delta_{\delta\eta})
\nonumber\\
&&{}\times \!\sum_{{\bf k'}\atop{{\bf k''}=\,{\bf k}-{\bf k'}}}
\int_0^t\!\int_0^t
\bbo_{\delta\eta}\,|\BoTbf_{\bf k'}|^2\,
\langle\ViT_{{\bf k''}\alpha}(t')\ViT^*_{{\bf k''}\beta}(t'')\rangle
\,dt'dt''.
\label{BiTS}
\eeq
Here, we use $\langle\ViT_{{\bf k''}\alpha}\ViT^*_{{\bf k^{iv}}\beta}\rangle\propto
\delta_{{\bf k''},{\bf k^{iv}}}$, see equation~(\ref{ViT_ViT_AVERAGED_NEW}), and 
therefore, ${\bf k^{iv}}={\bf k''}$ and ${\bf k'''}={\bf k'}$. We also assume that
$\bbo_{\alpha\beta}={\rm const}$ (our first working hypothesis), and therefore,
$\BoT_{{\bf k'}\delta}\BoT^*_{{\bf k'}\eta}=\bbo_{\delta\eta}|\BoT_{\bf k'}|^2=
\bbo_{\delta\eta}|\BoTbf_{\bf k'}|^2$. 

Now, according to formula~(\ref{M_K}), we need to integrate equation~(\ref{BiTS}) 
over all directions of the unit vector ${\bf\k}$. We carry out this integration in 
Appendix~\ref{COUPLING_TERMS_DERIVATION}, the result is
\beq
\int k^2\, \langle|\BiTbf_{\bf k}(t)|^2\rangle\,d^2{\bf\k} \!\!&=&\!\! t
\int_0^\infty\!dk'\:{\cal K}(k,k')\,\int\!k'^2\,|\BoTbf_{\bf k'}|^2\,d^2{\bf\k'},
\label{BiTS_TERM}
\eeq
\beq
K(k,k') \!\!&=&\!\! 
k^4 {\left(\frac{L}{2\pi}\right)}^{\!3}
\int_0^\pi d\theta\,\sin^3\theta\,J_{0k''} \int_0^{2\pi} d\varphi\, 
\left(1+{\bar\Omega}''/\Omega''_{\rm rd}\right)^2
\,\left\{\,\frac{k'^2+2k(k-k'\cos\theta)\cos^2\varphi}{k''^2}\right.
\nonumber\\
{}\!\!&-&\!\! \left.
\left[1-\left(1+2\Omega''/\Omega''_{\rm rd}\right)^{-2}\right]\,
\frac{k^2}{k''^2}\,\frac{(k'-k\cos\theta)^2+(k^2-k'^2)\sin^2\theta\sin^2\varphi}
{(k'-k\cos\theta)^2+k^2\sin^2\theta\sin^2\varphi}\cos^2\varphi\right\},
\qquad
\label{COUPLING_KERNEL}
\\
k'' &=& (k^2+k'^2-2kk'\cos\theta)^{1/2},
\\
{\bar\Omega}''/\Omega''_{\rm rd} &=& 6\,\frac{k''^2}{k_\nu^2},
\qquad
2\Omega''/\Omega''_{\rm rd} \:\;=\;\: 
90\,\frac{k^2}{k_\nu^2}\,\frac{(k'-k\cos\theta)^2+k^2\sin^2\theta\sin^2\varphi}{k''^2}
\sin^2\theta\cos^2\varphi,
\label{OMEGA_RATIO_FIRST}
\eeq
where function $J_{0k}$ is given by equation~(\ref{J_ZERO_K}).

Next, we calculate the 
$\langle\BiiT_{{\bf k}\alpha}(t)\rangle\BoT^*_{{\bf k}\alpha}+{\rm c.c.}$ term of 
equation~(\ref{BkS_EXPANSION}). First, we integrate equation~(\ref{B_EVOLUTION_ii}) 
in time, and then Fourier transform the result in space, 
${\bf r}\rightarrow{\bf k}$. We have
\beq
\BiiT_{{\bf k}\eta}(t) = ik_\tau(\delta_{\eta\beta}\delta_{\tau\chi}
-\delta_{\tau\beta}\delta_{\eta\chi})
\int_0^t \!\!\!\sum_{{\bf k'}\atop{{\bf k''}=\,{\bf k}-{\bf k'}}}\!\!\!
\!\left[\ViT_{{\bf k''}\beta}(t')\,\BiT_{{\bf k'}\chi}(t')
+\ViiT_{{\bf k''}\beta}(t')\,\BoT_{{\bf k'}\chi}\right] dt',
\quad
\label{BiiT}
\eeq
where we use again the divergence free conditions, $k_\alpha\BoT_{{\bf k}\alpha}=0$, 
$k_\alpha\BiT_{{\bf k}\alpha}=0$, $k_\alpha\ViT_{{\bf k}\alpha}=0$ and 
$k_\alpha\ViiT_{{\bf k}\alpha}=0$. 
Second, we ensemble average his equation. The second term in the brackets $[...]$ 
averages out because of equation~(\ref{Vii_AVERAGE_NEW}). Third, we multiply the 
averaged equation by $\BoT^*_{{\bf k}\eta}$, add the complex conjugate, and
use formula~(\ref{BiT}) for $\BiT_{{\bf k'}\chi}$. We have
\beq
\langle\BiiT_{{\bf k}\eta}\rangle\BoT^*_{{\bf k}\eta}+{\rm c.c.} 
\!\!&=&\!\! ik_\tau(\delta_{\eta\beta}\delta_{\tau\chi}
-\delta_{\tau\beta}\delta_{\eta\chi})\, i(\delta_{\chi\alpha}\delta_{\gamma\delta}
-\delta_{\gamma\alpha}\delta_{\chi\delta})
\nonumber\\
\!\!&&\!\!{}\times \sum_{{\bf k'}\atop{{\bf k''}=\,{\bf k}-{\bf k'}}}
\sum_{{\bf k'''}\atop{{\bf k^{iv}}=\,{\bf k'}-{\bf k'''}}} \!\!\!\!\! k'_\gamma
\int_0^t\!dt'\!\int_0^{t'}\!dt''\,
\langle\ViT_{{\bf k^{iv}}\alpha}(t'')\,\ViT_{{\bf k''}\beta}(t')\rangle\,
\BoT^*_{{\bf k}\eta}\BoT_{{\bf k'''}\delta} + {\rm c.c.}
\nonumber\\
{}\!\!&=&\!\!\! {}-k_\tau(\delta_{\alpha\tau}\delta_{\eta\beta}\delta_{\gamma\delta}
-\delta_{\alpha\eta}\delta_{\tau\beta}\delta_{\gamma\delta}
+\delta_{\delta\eta}\delta_{\gamma\alpha}\delta_{\tau\beta})\,
\BoT^*_{{\bf k}\eta}\BoT_{{\bf k}\delta}
\nonumber\\
\!\!&&\!\!{}\times \sum_{\bf k''} \,(k_\gamma-k''_\gamma) \int_0^t\!dt'\!\int_0^{t'}\!dt''\,
\langle\ViT_{-{\bf k''}\alpha}(t'')\,\ViT_{{\bf k''}\beta}(t')\rangle + {\rm c.c.}
\nonumber\\
{}\!\!&=&\!\!\! {}-2k_\alpha k_\beta\,|\BoTbf_{\bf k}|^2
\sum_{\bf k''} \int_0^t\!dt'\!\int_0^{t'}\!dt''\,
\langle\ViT_{-{\bf k''}\alpha}(t'')\,\ViT_{{\bf k''}\beta}(t')\rangle
\nonumber\\
{}\!\!&+&\!\! 2k_\tau(\delta_{\alpha\tau}\delta_{\eta\beta}
-\delta_{\alpha\eta}\delta_{\tau\beta})\,\bbo_{\eta\gamma}\,|\BoT_{\bf k}|^2
\sum_{\bf k''} \,k''_\gamma \int_0^t\!dt'\!\int_0^{t'}\!dt''\,
\langle\ViT_{-{\bf k''}\alpha}(t'')\,\ViT_{{\bf k''}\beta}(t')\rangle
\nonumber\\
{}\!\!&=&\!\!\! {}-2k^2|\BoTbf_{\bf k}|^2\,{\left(\frac{L}{2\pi}\right)}^{\!3}\!
\int_{-\infty}^\infty \!d^3{\bf k''}\! \int_0^t\!dt' \int_0^{t'}\!dt''\:
\k_\alpha\langle\ViT_{-{\bf k''}\alpha}(t'')\,\ViT_{{\bf k''}\beta}(t')\rangle\k_\beta .
\quad\;
\label{BiiT_AVERAGED}
\eeq
Here, we use $\langle\ViT_{{\bf k^{iv}}\alpha}\ViT_{{\bf k''}\beta}\rangle\propto
\delta_{{\bf k^{iv}},-{\bf k''}}$, see equation~(\ref{ViT_ViT_AVERAGED_NEW}), and 
therefore, ${\bf k^{iv}}=-{\bf k''}$ and ${\bf k'''}={\bf k}$. We use the field 
divergence free condition, $k_\alpha\BoT_{{\bf k}\alpha}=0$, this is why there are 
only three terms in the third line of equation~(\ref{BiiT_AVERAGED}). On the fourth 
line of equation~(\ref{BiiT_AVERAGED}) we use ${\bf k'}={\bf k}-{\bf k''}$ to change
the summation over ${\bf k'}$ to the summation over ${\bf k''}$. To obtain the fifth 
and sixth lines of equation~(\ref{BiiT_AVERAGED}), we use 
$k_\alpha\BoTbf_{{\bf k}\alpha}=0$, $k''_\alpha\ViT_{-{\bf k''}\alpha}=0$ and 
$\BoT_{{\bf k}\eta}\BoT^*_{{\bf k}\delta}=\bbo_{\eta\delta}|\BoT_{\bf k}|^2$
(because $\bbo_{\eta\delta}={\rm const}$). The two terms in the sixth line of 
equation~(\ref{BiiT_AVERAGED}) cancel each other because of the symmetry of tensor 
$\langle\ViT_{-{\bf k''}\alpha}\,\ViT_{{\bf k''}\beta}\rangle$ with respect to 
the exchange $\alpha\leftrightarrow\beta$ [see eq.~(\ref{ViT_ViT_AVERAGED_NEW})]. 
We change the summation over ${\bf k''}$ to the integration over ${\bf k''}$ in 
the last line of equation~(\ref{BiiT_AVERAGED}).

Note that ${\bf k}$ is perpendicular to $\bobf$ because the field is divergence free, 
$k_\alpha\BoT_{{\bf k}\alpha}=0$. Therefore, to carry out the integration over 
${\bf k''}$ in equation~(\ref{BiiT_AVERAGED}), we use a spherical 
system of coordinates, $d^3{\bf k''} = k''^2dk''\sin\theta\,d\theta\,d\varphi$. 
Here $\theta$ is the angle between ${\bf k''}$ and ${\bf k}$, so that 
${\bf\k''}\cdot{\bf\k} = \cos\theta$, and $\varphi$ is the angle between $\bobf$ 
and the projection of ${\bf k''}$ on the plane perpendicular to ${\bf k}$, so that 
$\mu'' = {\bf\k''}\cdot\bobf = \sin\theta\cos\varphi$.~\footnote{
Note that these angles $\theta$ and $\varphi$ have no any relation to the
angles in equation~(\ref{COUPLING_KERNEL}).
}
Using equation~(\ref{ViT_ViT_AVERAGED_NEW}), and equation~(\ref{H_F_INTEGRAL_T'T}) 
in the limit $t\gg\tau$, we obtain
\beq
&&\int_0^t\!dt'\int_0^{t'}\!dt''\;\k_\alpha
\langle\ViT_{-{\bf k''}\alpha}(t'')\,\ViT_{{\bf k''}\beta}(t')\rangle\k_\beta
\nonumber\\
&&\qquad {}=\frac{1}{2} J_{0k''}t \left(1+{\bar\Omega}''/\Omega''_{\rm rd}\right)^2
\sin^2\theta \left\{ 1-\left[1-\left(1+2\Omega''/\Omega''_{\rm rd}\right)^{-2}\right]
\frac{\cos^2\theta\cos^2\varphi}{\cos^2\theta\cos^2\varphi+\sin^2\varphi}\right\}.
\qquad
\label{TT'_INTEGRAL_FOR_BiiT}
\eeq
Now, we substitute this equation and formula 
$d^3{\bf k''} = k''^2dk''\sin\theta\,d\theta\,d\varphi$ into 
equation~(\ref{BiiT_AVERAGED}), and integrate the result over all directions of 
the unit vector ${\bf\k}$. We have
\beq
&&\int\!k^2 \left[\langle\BiiT_{{\bf k}\eta}\rangle\BoT^*_{{\bf k}\eta}+{\rm c.c.}\right]
d^2{\bf\k} = {}- t\,k^2 \int k^2\,|\BoTbf_{\bf k}|^2\,d^2{\bf\k}
\;\;{\left(\frac{L}{2\pi}\right)}^{\!3}\!
\int_0^\infty\!k''^2J_{0k''}\,dk'' \int_0^\pi\!\sin^3\theta\,d\theta 
\nonumber\\
&&\qquad\quad {}\times \int_0^{2\pi}\!d\varphi\, 
\left(1+{\bar\Omega}''/\Omega''_{\rm rd}\right)^2
\left\{ 1-\left[1-\left(1+2\Omega''/\Omega''_{\rm rd}\right)^{-2}\right] 
\frac{\cos^2\theta\cos^2\varphi}{\cos^2\theta\cos^2\varphi+\sin^2\varphi}\right\}.
\label{BiiT_TERM}
\eeq
Here, ${\bar\Omega}''$, $\Omega''$ and $\Omega''_{\rm rd}$ depend on $k''$ and on 
$\mu''^2=\sin\theta\cos\varphi$, see equations~(\ref{OMEGA}) and~(\ref{OMEGA_DAMP}).

Finally, we substitute equations~(\ref{BkS_EXPANSION}),~(\ref{BiTS_TERM}) 
and~(\ref{BiiT_TERM}) into equation~(\ref{M_K}) for the magnetic energy spectrum 
$M(t, k)$. We choose $t$ small enough for the quasi-linear expansion to be valid, 
so that $\partial_t\langle M(t, k)\rangle=[M(t, k)-M(0, k)]/t$. As a result, we 
obtain {\it the mode coupling equation} for the magnetic energy spectrum,
\beq
\frac{\partial M}{\partial t} = \int_0^\infty K(k,k')M(t,k')\,dk' -
2\frac{\eta_{\mbox{\tiny$T$}}}{4\pi}\,k^2 M(t,k),
\label{MODE_COUPLING}
\eeq
where the mode coupling kernel $K(k,k')$ is given by 
equations~(\ref{COUPLING_KERNEL})--(\ref{OMEGA_RATIO_FIRST}),
and the turbulent diffusion constant
\beq
\frac{\eta_{\mbox{\tiny$T$}}}{4\pi} &=& \frac{1}{2}{\left(\frac{L}{2\pi}\right)}^{\!3}
\int_0^\infty k'''^2J_{0k'''}\,dk'''\int_0^\pi\sin^3\theta\:d\theta\int_0^{2\pi}d\varphi\,
\left(1+{\bar\Omega}'''/\Omega'''_{\rm rd}\right)^2
\nonumber\\
&&\quad{}\times \left\{1-\left[1-\left(1+2\Omega'''/\Omega'''_{\rm rd}\right)^{-2}\right]
\frac{\cos^2\theta\cos^2\varphi}{\cos^2\theta\cos^2\varphi+\sin^2\varphi}\right\},
\label{ETA_T}
\\
{\bar\Omega}'''/\Omega'''_{\rm rd} &=& 6\,\frac{k'''^2}{k_\nu^2},
\qquad
2\Omega'''/\Omega'''_{\rm rd} \:\;=\;\:
90\,\frac{k'''^2}{k_\nu^2}\,\sin^2\theta\cos^2\varphi\,(1-\sin^2\theta\cos^2\varphi).
\label{OMEGA_RATIO_SECOND}
\eeq
The function $J_{0k}$ is given by equation~(\ref{J_ZERO_K}). Here, we use 
equations~(\ref{OMEGA}) and~(\ref{OMEGA_DAMP}). To obtain 
equation~(\ref{OMEGA_RATIO_FIRST}) we use equations~(\ref{MU''}). To obtain 
equation~(\ref{OMEGA_RATIO_SECOND}), we use formula $\mu''^2=\sin\theta\cos\varphi$. 
In the last three equations for the turbulent diffusion constant we replace $k''$ by 
$k'''$ in order to distinguish it from $k''$ in the equations~(\ref{COUPLING_KERNEL}) 
for the coupling kernel. Note, that according to the mode coupling 
equation~(\ref{MODE_COUPLING}), the $k''$ modes of the turbulence interact with 
the $k'$ modes of the magnetic field to change the energy in the $k$ modes of the 
magnetic field.

In the kinematic turbulent dynamo case, we take the limit 
$\Omega_{\rm rd}\rightarrow \infty$. In this limit, as one might expect, after 
integrating over $\varphi$, equation~(\ref{COUPLING_KERNEL}) reduces to the 
mode coupling equation of Kulsrud and Anderson (1992). Equation~(\ref{ETA_T}) 
for the turbulent diffusion constant, after integrating over both $\theta$ and 
$\varphi$, reduces to the corresponding equation of Kulsrud and Anderson (1992)
[see also Kraichnan and Nagarajan (1967), who obtained the same equation in a
different form].

\subsection
{The Magnetic Energy Spectrum on Subviscous Scales}
\label{SMALL_SCALES}

Equation~(\ref{MODE_COUPLING}), which gives the evolution of the magnetic 
energy spectrum in the magnetized turbulent dynamo theory, is the principal 
result of this paper. However, this equation is rather complicated for 
easy interpretation. In this section we limit ourselves to the evolution 
of the magnetic energy spectrum on small subviscous scales, which is of great 
interest during the magnetized dynamo stage in a 
protogalaxy~\footnote{
The time evolution of the magnetic energy on large scales is slow because it 
is set by the turnover time of large turbulent eddies. As a result, the 
large-scale field is not of significant interest until the small-scale field 
saturates, the magnetized dynamo stage ends and the inverse cascade stage begins.
}.
In this limit, $k\gg k_\nu$, and the integro-differential 
equation~(\ref{MODE_COUPLING}) for the magnetic spectrum simplifies to an 
ordinary differential equation. 

Let us refer to equation~(\ref{COUPLING_KERNEL}) for the mode coupling kernel 
$K(k,k')$. Function $J_{0k''}$ cuts off at the viscous wave number $k_\nu$
[see eq.~(\ref{J_ZERO_K})]. Therefore, in the large-$k$ limit, $k\gg k_\nu$, 
we have $k''\sim|k-k'|\ll k,k'$, and can expand the kernel $K(k,k')$. However, 
the simplest way of calculations is to introduce an arbitrary function of $k$, 
$F(k)$, which varies slowly in the region $k\gg k_\nu$, and vanishes outside 
of this region (Kulsrud \& Anderson 1992). To derive the mode coupling 
equation on small (subviscous) scales, we calculate the following integral
\beq
&&\!\!\!\!\!
\int_0^\infty\!F(k)\frac{\partial M}{\partial t}\,dk =
\frac{1}{4\pi\rho}{\left(\frac{L}{2\pi}\right)}^{\!3}\!\!\int_{-\infty}^\infty\! F
\frac{\partial\langle|{\bf{\tilde B}}_{\bf k}|^2\rangle}{\partial t}\,d^3{\bf k}
=\frac{1}{4\pi\rho}{\left(\frac{L}{2\pi}\right)}^{\!3}\!\!\int_{-\infty}^\infty\!
F \frac{\langle|{\bf{\tilde B}}_{\bf k}(t)|^2\rangle
-|{\bf{\tilde B}}_{\bf k}(0)|^2}{t}\,d^3{\bf k}
\nonumber\\
&&\qquad{}=\frac{1}{4\pi\rho}{\left(\frac{L}{2\pi}\right)}^{\!3}\,\frac{1}{t} \left\{\,
\int_{-\infty}^\infty\! F(k)\,\langle|\BiTbf_{\bf k}(t)|^2\rangle\,d^3{\bf k}
+\int_{-\infty}^\infty\! F(k)
\Big[\langle\BiiT_{{\bf k}\alpha}(t)\rangle\BoT^*_{{\bf k}\alpha}+{\rm c.c.}\Big]
\,d^3{\bf k}\right\}
\nonumber\\
&&\qquad{}= \frac{\Gamma}{5}\:\int_0^\infty F(k) \left[k^2\frac{\partial^2M}{\partial k^2}
-(\Lambda_1-1)k\frac{\partial M}{\partial k} + \Lambda_0 M\right] dk.
\label{F_M_INTEGRAL}
\eeq
Here, to obtain the first line, we use equation~(\ref{M_K}) and replace the time 
derivative by the time finite difference, assuming that $t$ is small, and our 
quasi-linear expansion is valid. To obtain the second line, we use 
equation~(\ref{BkS_EXPANSION}). The derivation of the final result [the 3rd line] 
is given in Appendix~\ref{SMALL_SCALES_DERIVATION}. The constants $\Gamma$, 
$\Lambda_1$ and $\Lambda_0$ are 
\beq
\Gamma \!\!&=&\!\! \frac{5}{2}{\left(\frac{L}{2\pi}\right)}^{\!3}
\int_0^\infty k^4J_{0k}\,dk \int_0^\pi d\theta\, \sin^3\theta\cos^2\theta
\nonumber\\
\!\!&&\!\! {}\times\int_0^{2\pi} d\varphi\,\left(1+{\bar\Omega}/\Omega_{\rm rd}\right)^2\!
\left\{ 1-\left[1-\left(1+2\Omega/\Omega_{\rm rd}\right)^{-2}\right] 
\frac{\cos^2\theta\cos^2\varphi}{\cos^2\theta\cos^2\varphi+\sin^2\varphi}\right\}\!,
\qquad\quad
\label{CAPITAL_GAMMA}
\\
\Lambda_1 \!\!&=&\!\! -3 +\frac{5}{\Gamma}{\left(\frac{L}{2\pi}\right)}^{\!3}
\int_0^\infty k^4 J_{0k}\,dk \int_0^\pi\! d\theta\, \sin^3\theta
\int_0^{2\pi}\!\! d\varphi \left(1+{\bar\Omega}/\Omega_{\rm rd}\right)^2\,
\left\{ 2\cos^2\theta\cos^2\varphi +\frac{1}{2}\sin^2\theta \right.
\nonumber\\
\!\!&&\!\! {}-\left. \left[1-\left(1+2\Omega/\Omega_{\rm rd}\right)^{-2}\right] 
\left(2+\frac{1}{2}\,\frac{\sin^2\theta}{\cos^2\theta\cos^2\varphi+\sin^2\varphi}\right)
\cos^2\theta\cos^2\varphi \right\}\!,
\label{LAMBDA_1}
\\
\Lambda_0 \!\!&=&\!\! 2 + \frac{5}{\Gamma}{\left(\frac{L}{2\pi}\right)}^{\!3}
\int_0^\infty k^4J_{0k}\,dk \int_0^\pi d\theta\, \sin^3\theta \int_0^{2\pi} \!\!d\varphi 
\left(1+{\bar\Omega}/\Omega_{\rm rd}\right)^2\,
\left\{2\sin^2\theta\cos^2\varphi -\frac{1}{2}\sin^2\theta \right.
\nonumber\\
\!\!&&\!\! {}+\left. \left[1-\left(1+2\Omega/\Omega_{\rm rd}\right)^{-2}\right] 
\left(\cos^2\theta-\sin^2\theta+\frac{1}{2}\,
\frac{\sin^2\theta\cos^2\theta}{\cos^2\theta\cos^2\varphi+\sin^2\varphi}
\right)\cos^2\varphi \right\}\!,
\label{LAMBDA_0}
\\
{\bar\Omega}/\Omega_{\rm rd} \!\!\!&=&\!\! 6\,\frac{k^2}{k_\nu^2}\,,
\qquad
2\Omega/\Omega_{\rm rd} \:=\:
90\,\frac{k^2}{k_\nu^2}\,\sin^2\theta\cos^2\varphi\,(1-\sin^2\theta\cos^2\varphi),
\label{OMEGA_RATIO_THIRD}
\eeq
and function $J_{0k}$ is given by equation~(\ref{J_ZERO_K}).

Equation~(\ref{F_M_INTEGRAL}) is valid for an arbitrary function $F$. As a result, 
the integrands in the left- and right-hand-side of this equation should be equal, 
and we finally obtain the mode-coupling equation for the magnetic energy spectrum 
$M(t,k)$ on small (subviscous) scales
\beq
\frac{\partial M}{\partial t} = 
\frac{\Gamma}{5} \left[k^2\frac{\partial^2M}{\partial k^2}
-(\Lambda_1-1)k\frac{\partial M}{\partial k} + \Lambda_0 M\right].
\label{SMALL_SCALES_MODE_COUPLING}
\eeq
The constant $\Gamma$ and the constant dimensionless numbers $\Lambda_0$ and $\Lambda_1$ 
can easily be calculated numerically. The result is (Malyshkin 2001)
\beq
&&\Gamma \approx 11\left(U_0L/\nu\right)^{1/2}\,(U_0/L),
\qquad
\Lambda_1 \approx 2,
\qquad
\Lambda_0 \approx 5,
\label{GAMMA_LAMBDA_RESULT}
\eeq
and therefore,
\beq
&&\frac{\partial M}{\partial t} = 2.2\left(\frac{U_0L}{\nu}\right)^{1/2}\frac{U_0}{L} 
\left[k^2\frac{\partial^2M}{\partial k^2}-k\frac{\partial M}{\partial k} + 5M\right].
\label{SMALL_SCALES_MODE_COUPLING_RESULT}
\eeq
(It is easy to see that $\Lambda_1=2$ and $\Lambda_0=5$ correspond to the limit 
$\Omega_{\rm rd}\ll\nu k_\nu^2\,$.)

If for the moment we consider the kinematic turbulent dynamo, then we take the limit 
$\Omega_{\rm rd}\rightarrow \infty$ in equations~(\ref{CAPITAL_GAMMA})--(\ref{LAMBDA_0}). 
After integrating over $\varphi$ and $\theta$, we have $\Gamma=\gamma_{\rm o}$, where 
$\gamma_{\rm o}$ is the Kulsrud-Anderson (1992) magnetic energy growth rate, 
$\Lambda_1=3$ and $\Lambda_0=6$. In this case equation~(\ref{SMALL_SCALES_MODE_COUPLING}) 
coincides with the corresponding equation of Kulsrud and Anderson (1992) and with the 
corresponding equation of Kazantsev (1968), as one might expect (note that Kazantsev's
equation is given in the Fourier space).

Now, assume that $M(t,k_{\rm ref})$ is known as a function of time at some reference 
wave number $k=k_{\rm ref}$. Then the solution of~(\ref{SMALL_SCALES_MODE_COUPLING}) is
\beq
M(t,k)=\int_{-\infty}^t M(t',k_{\rm ref}) G(k/k_{\rm ref},t-t')\,dt',
\label{SMALL_SCALES_SOLUTION}
\eeq
where the Green's function $G(k,t)$ is
\beq
G(k,t) = \sqrt{\frac{5}{4\pi}}\: \frac{k^{\Lambda_1/2}\ln{k}}{\Gamma^{1/2}t^{3/2}}
\,e^{(\Gamma/5)(\Lambda_0-\Lambda_1^2/4)t} \,e^{\left.-5\ln^2{k}\right/4\Gamma t}
= \sqrt{\frac{5}{4\pi}}\: \frac{k\ln{k}}{\Gamma^{1/2}t^{3/2}}
\,e^{(4\Gamma/5)t}\, e^{\left.-5\ln^2{k}\right/4\Gamma t}.
\label{GREENS_FUNCTION}
\eeq
[This Green function can be obtained by applying the Laplace transformation in 
time, $t\rightarrow s$, to equation~(\ref{SMALL_SCALES_MODE_COUPLING}), see
Kulsrud \& Anderson 1992; Malyshkin 2001]. To obtain the final result in 
equation~(\ref{GREENS_FUNCTION}), we use equations~(\ref{GAMMA_LAMBDA_RESULT}) 
for $\Lambda_1$ and $\Lambda_0$. 

According to equations~(\ref{SMALL_SCALES_SOLUTION}) and~(\ref{GREENS_FUNCTION}), 
we see that a ``signal'' $M(t,k_{\rm ref})$, at zero time, will increase 
exponentially as $e^{(4/5)\Gamma t}$ and will extend down to the scale 
$k_{\rm peak}\approx e^{(4/5)\Gamma t}\,k_{\rm ref}$, where $k_{\rm peak}$ is 
the peak of function $kG(k,t)$, (of course, the field scale can not become
less than the resistivity scale). As a result, in the magnetized dynamo theory
the magnetic energy tends to quickly propagate to very small subviscous scales,
the same way as it does in the kinematic dynamo theory (Kulsrud \& Anderson 1992;
Schekochihin, Boldyrev, \& Kulsrud 2002). However, the tail of the magnetic energy spectrum 
on $k_{\rm ref}\simlt k\simlt k_{\rm peak}$ scales increases with the wave number 
as ${}\propto {\it k}^{\Lambda_1/2}={\it k}$ instead of 
${}\propto {\it k}^{\rm 3/2}$ in the kinematic theory (Kulsrud and Anderson 1992). 
Note, that according to equations~(\ref{GAMMA_RESULT}) and~(\ref{GAMMA_LAMBDA_RESULT}), 
the growth rate of the Green's function, $(4/5)\Gamma$, is approximately equal to a 
half of the growth rate of the total magnetic energy (the latter one is $2\gamma$).

\section
{Discussion and Conclusions}
\label{DISCUSSION}

In this paper we have developed a theoretical basis for the magnetized 
turbulent dynamo, which operates in protogalaxies, where the plasma is fully 
ionized, and the viscosity is the Braginskii tensor viscosity. The results of 
the kinematic dynamo theory, already seem to support the primordial 
(protogalactic) dynamo origin of cosmic magnetic fields (Kulsrud \etal 1997). 
The results that we have obtained for the magnetized dynamo, further support 
this primordial origin theory. This is because the number of the magnetic energy
e-foldings during the magnetized turbulent dynamo stage in a protogalaxy, given by 
equation~(\ref{GAMMA_COLLAPSE_TIME}), is up to ten time larger than that in the 
kinematic dynamo theory. This number of e-foldings is more than large enough for 
the magnetic field in the protogalaxy to grow from its seed value, provided by the 
Biermann battery, up to the field-turbulence energy equipartition value. The number 
of e-foldings of the magnetic energy on the viscous scale, which is determined by 
the growth rate $(4/5)\Gamma$ of the Green's function~(\ref{GREENS_FUNCTION}), is 
less by one half, but it is still sufficiently 
large~\footnote{
Of course, our results~(\ref{GAMMA})--(\ref{GAMMA_RATIO}) for the magnetic 
energy growth rate are sensitive to the value of the physical parameter 
$\Omega_{\rm rd}$, which is estimated in equation~(\ref{OMEGA_DAMP}). We 
also left out the finite time correlation effects. Therefore, our result 
for the number of magnetic energy e-foldings should be viewed as an estimate, 
valid within a factor of order two. However, it is important that the number 
of e-foldings, which we found, is a very large number.
}. 

Another our prediction is that the tail of the magnetic energy 
spectrum on small subviscous scales increases with the wave number as 
${}\propto {\it k}$ [see the Green's function~(\ref{GREENS_FUNCTION})], 
instead of ${}\propto {\it k}^{\rm 3/2}$ in the kinematic 
theory~\footnote{
This our prediction is not sensitive to the value of $\Omega_{\rm rd}$,
as long as $\Omega_{\rm rd}\ll \nu k_\nu^2$.
}.
Therefore, in the magnetized dynamo the magnetic energy is 
slightly shifted to larger scales as compared to the kinematic dynamo 
case.

The Green's function solution~(\ref{GREENS_FUNCTION}) indicates that 
in the magnetized dynamo theory the magnetic energy tends to quickly propagate to 
very small subviscous scales, similar to the kinematic dynamo case. On the 
other hand, the observed cosmic fields have rather large correlation lengths. 
Therefore, the magnetic field lines must be unwrapped on small scales by 
the Lorentz tension forces, while the field energy is transferred and 
amplified on larger scales during the inverse cascade stage. This most important 
and most interesting stage happens when the magnetic field energy is comparable 
to the turbulent kinetic energy. In the final part of this paper let us discuss the 
significance of the Braginskii viscosity for the inverse cascade in more details.

First, note that the theory of the inverse cascade in a plasma with the regular 
isotropic (non-Braginskii) viscosity has a difficulty of unwrapping of the 
small-scale magnetic field lines. This difficulty can be understood as 
follows~\footnote{
Cowley, Kulsrud, \& Schekochihin 2001, private communications.
}.
The equation for the turbulent velocities ${\bf V}$ in the plasma with the 
isotropic viscosity, including the Lorentz forces, is (Landau \& Lifshitz 1984)
\beq
\partial_t V_\alpha &=&
-P_{,\alpha}+f_\alpha+\nu\triangle V_\alpha
+(1/4\pi\rho)({\bf B}\cdot{\bf\nabla}) B_\alpha
-({\bf V}\cdot{\bf\nabla})V_\alpha
\label{EQUATION_FOR_U_LORENTS_FORCES}
\eeq
[compare with eq.~(\ref{EQUATION_FOR_U})]. We can estimate the unwrapping velocity, 
$V_{\rm unwrap}$, by Fourier transforming 
equation~(\ref{EQUATION_FOR_U_LORENTS_FORCES}) in space, ${\bf r}\rightarrow{\bf k}$, 
and then balancing the viscous and the inertial forces against the magnetic tension 
force. The isotropic viscosity dominates on small scales, and we have
\beq
\nu k^2 V_{\rm unwrap} \sim (1/4\pi\rho)k_\parallel B^2,
\qquad
V_{\rm unwrap} \sim (k_\parallel/k)\,(k_\nu/k)\,(V_{\rm A}^2/\nu k_\nu) \ll V_{\rm A},
\label{UNWRAPPING_VELOCITY}
\eeq
where $V_{\rm A}$ is the Alfven speed. At the time of the field-turbulence energy 
equipartition on the viscous scale, the Alfven speed is $V_{\rm A}\sim\nu k_\nu$, 
and before this equipartition $V_{\rm A}<\nu k_\nu$. Since in the kinematic dynamo 
theory the field lines have a folding pattern, $k\gg k_\parallel\sim k_\nu$ (see 
Fig.~\ref{FIGURE_FOLDING}), the unwrapping velocity~(\ref{UNWRAPPING_VELOCITY}) is 
small compared to the Alfven speed, even at the equipartition time. In other words, 
since $V_{\rm A}\sim V$ at the energy equipartition, then 
$k_\parallel V_{\rm A}\sim\gamma$ (here $2\gamma$ is the magnetic energy growth rate), 
and the unwrapping rate, $k_\parallel V_{\rm unwrap}$, is much smaller than $\gamma$. 
This means that the field continues to grow on the viscous and subviscous scales even 
beyond the energy equipartition. 

However, in the case of the magnetized turbulent dynamo the viscosity term in 
equation~(\ref{EQUATION_FOR_U_LORENTS_FORCES}) is modified, and the anti-unwrapping 
argument does not apply. Indeed, the field unwrapping velocity is parallel and varies 
in the direction perpendicular to the magnetic field lines. The large velocity gradient
perpendicular to the field lines, which leads to a large perpendicular
stress in the isotropic viscosity case, is suppressed in the case of the 
Braginskii viscous forces (because the transfer of the ion momentum in the 
perpendicular direction is inhibited). Therefore, in the magnetized dynamo theory
$k_\parallel V_{\rm unwrap}\sim k_\parallel V_{\rm A}\sim \gamma$ at the 
equipartition, and the magnetic field strength saturates on the viscous and 
subviscous scales. As a result, the Braginskii viscosity makes the inverse 
cascade of the magnetic energy more likely, because the larger turbulent eddies 
do not need to deliver their energy to the field on the viscous and subviscous 
scales~(Kulsrud 2000).

\acknowledgments
We are very grateful to Eric Blackman, Fausto Cattaneo, Steven Cowley, 
Bruce Draine, Jeremy Goodman, Robert Rosner, Alexander Schekochihin, 
David Spergel and Samuel Vainshtein for very useful and stimulating 
discussions and for a number of important comments. LM would like to 
thank the Department of Astrophysical Sciences at Princeton University 
for financial support. This work was partially supported by the DOE under 
the ASCI program at the University of Chicago and under DOE Contract No.
DE-AC 02-76-CHO-3073.

\appendix

\section{Details of the derivation of the mode coupling equation}
\label{COUPLING_TERMS_DERIVATION}

\begin{figure}[!t]
\vspace{7.0truecm}
\includegraphics{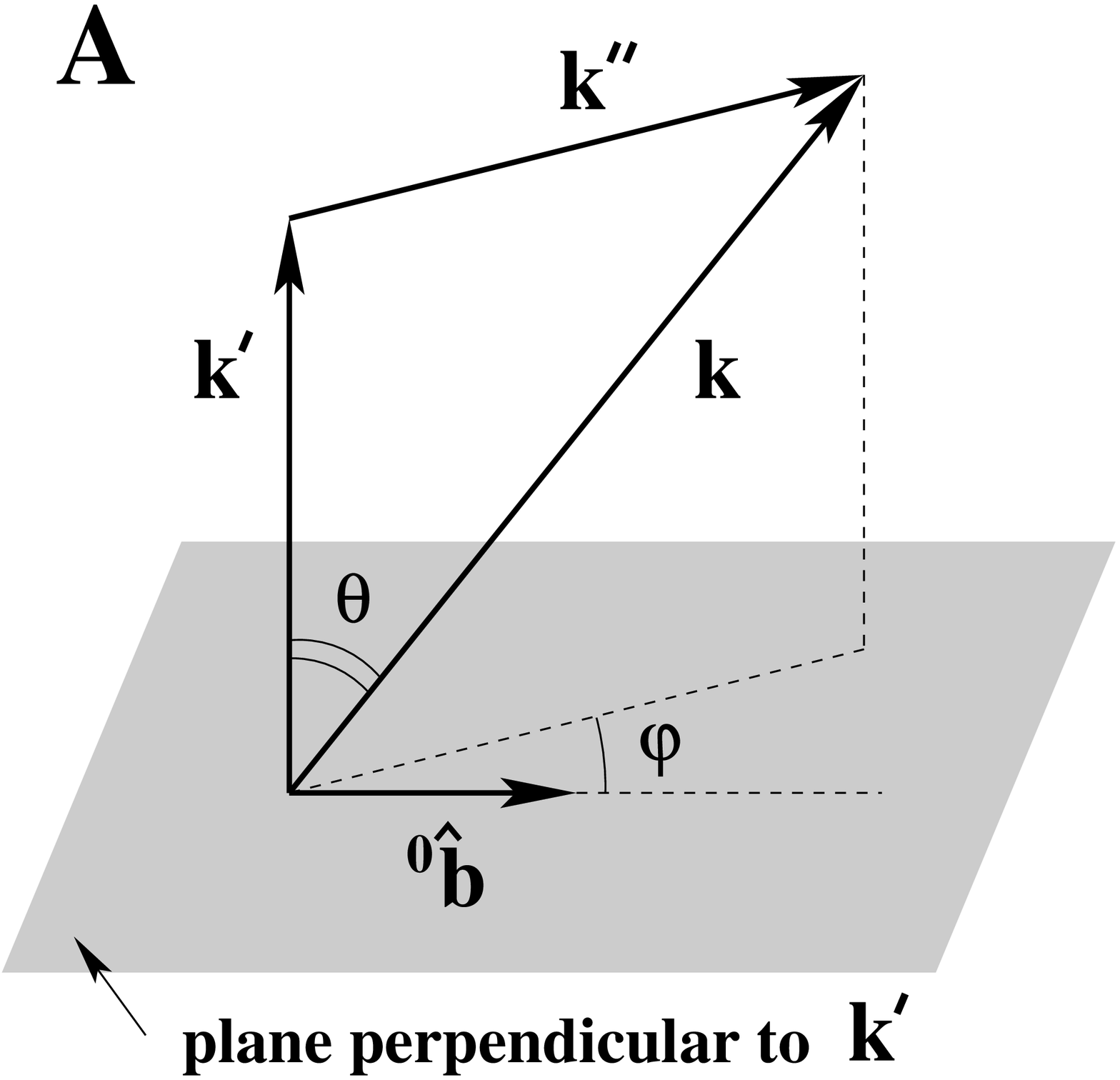}
\includegraphics{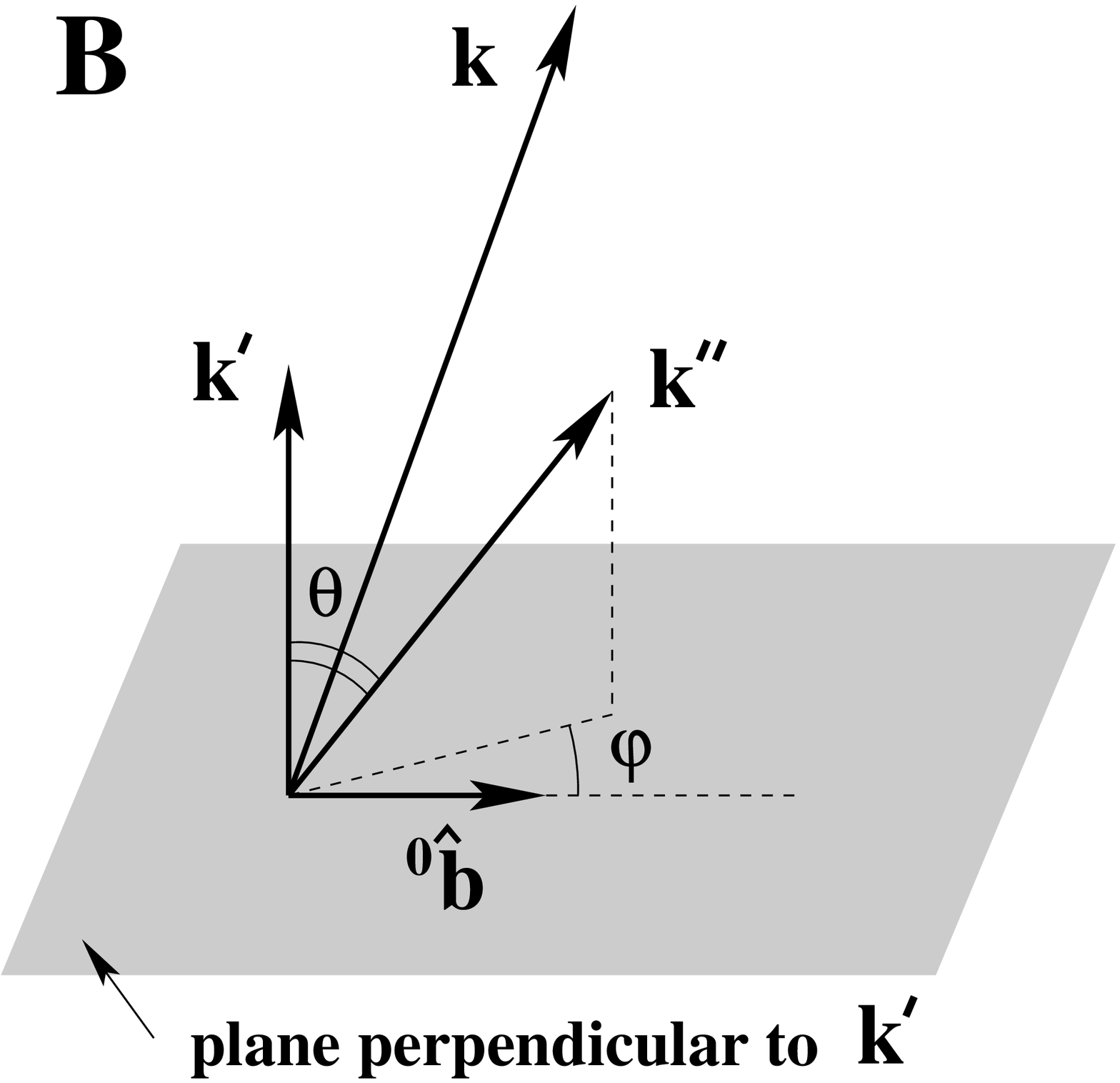}
\caption{
Relative position of vectors $\bobf$, ${\bf k'}$, ${\bf k''}$ and 
${\bf k}={\bf k'}+{\bf k''}$ in space. Note, that ${\bf k'}$ is
perpendicular to $\bobf$ because the magnetic field is divergence 
free. The plot on the left (A) refers to 
Appendix~\ref{COUPLING_TERMS_DERIVATION}, the plot on the right (B)
refers to Appendix~\ref{SMALL_SCALES_DERIVATION}.
}
\label{FIGURE_bo_and_k}
\end{figure}

Equation~(\ref{BiTS}) has four terms in the right-hand-side. Therefore, we have 
\beq
\int k^2\,\langle|\BiTbf_{\bf k}(t)|^2\rangle\,d^2{\bf\k} &=&
{\cal T}-{\cal T}'-{\cal T}''+{\cal T}''',
\label{BiTS_INTEGRATED_FOUR_TERMS}
\eeq
where there are also four terms,
\beq
{\cal T} \!\!\!\!&=&\!\!\! k^4 {\left(\frac{L}{2\pi}\right)}^{\!3}\!
\int_0^\infty\!dk'\!\!\int\!k'^2\,|\BoT_{\bf k'}|^2 d^2{\bf\k'}\!
\int\!\mu^2 d^2{\bf\k}\!\int_0^t\!\!\int_0^t 
\langle\ViT_{{\bf k''}\alpha}(t')\ViT^*_{{\bf k''}\alpha}(t'')\rangle\,dt'dt'',
\label{T}
\\
{\cal T}' \!\!\!\!&=&\!\!\! k^4 {\left(\frac{L}{2\pi}\right)}^{\!3}\!
\int_0^\infty\!dk'\!\!\int\!k'^2\,|\BoT_{\bf k'}|^2 d^2{\bf\k'}\!\int\!\mu\,d^2{\bf\k}\!
\int_0^t\!\!\int_0^t 
\bo_\alpha\langle\ViT_{{\bf k''}\alpha}(t')\ViT^*_{{\bf k''}\beta}(t'')\rangle\k_\beta
\,dt'dt'',
\label{T'}
\\
{\cal T}'' \!\!\!\!&=&\!\!\! {{\cal T}'}^* \!= {\cal T}',
\label{T''}
\\
{\cal T}''' \!\!\!\!&=&\!\!\! k^4 {\left(\frac{L}{2\pi}\right)}^{\!3}\!
\int_0^\infty\!dk'\!\!\int\!k'^2\,|\BoT_{\bf k'}|^2 d^2{\bf\k'}\!\int\!d^2{\bf\k}\!
\int_0^t\!\!\int_0^t 
\k_\alpha\langle\ViT_{{\bf k''}\alpha}(t')\ViT^*_{{\bf k''}\beta}(t'')\rangle\k_\beta
\,dt'dt'',
\label{T'''}
\eeq
Here, ${\bf k''}={\bf k}-{\bf k'}$, $\mu=({\bf\k}\cdot\bobf)$ [see eq.~(\ref{MU})], 
and we replace the summation over ${\bf k'}$ by the double integration over $k'$ and 
${\bf\k'}$. Next, refer to Figure~\ref{FIGURE_bo_and_k}A. Vector ${\bf k'}$ is 
perpendicular to $\bobf$ because the field is divergence free, 
$k'_\alpha\BoT_{{\bf k'}\alpha}=0$. The following useful equations are valid
(Malyshkin 2001),
\beq
&&{\bf\k}\cdot{\bf\k'}=\cos\theta\,,
\qquad
{\bf\k}\cdot{\bf\k''} = (k-k'\cos\theta)/k'',
\qquad
k''=(k^2+k'^2-2kk'\cos\theta)^{1/2},
\qquad
\label{kk'k''}
\\
&&\mu = {\bf\k}\cdot\bobf = \sin\theta\,\cos\varphi\,,
\qquad 
d^2{\bf\k} = \sin\theta\;d\theta\:d\varphi\,,
\label{MU_dk}
\\
&&\mu'' = \bobf\cdot{\bf\k''} = \frac{k}{k''}\,\sin\theta\cos\varphi\,,
\qquad
1-\mu''^2 = \frac{(k'-k\cos\theta)^2+k^2\sin^2\theta\sin^2\varphi}{k''^2}.
\label{MU''}
\eeq
Using these formulas, equation~(\ref{ViT_ViT_AVERAGED_NEW}) and 
equation~(\ref{H_F_INTEGRAL_TT}) in the limit $t\gg\tau$, it is straightforward 
to calculate the double time integral terms in equations~(\ref{T})--(\ref{T'''}),
\beq
\int_0^t\!\!\int_0^t\!
\langle\ViT_{{\bf k''}\alpha}(t')\ViT^*_{{\bf k''}\alpha}(t'')\rangle\,dt'dt'' 
\!\!\!&=&\!\!\! J_{0k''}t\left(1+{\bar\Omega}''/\Omega''_{\rm rd}\right)^2\,
\left[1+\left(1+2\Omega''/\Omega''_{\rm rd}\right)^{-2}\right],
\qquad\quad\!
\label{FIRST_DOUBLE_TIME_INTEGRAL}
\\
\int_0^t\!\!\int_0^t\!
\bo_\alpha\langle\ViT_{{\bf k''}\alpha}(t')\ViT^*_{{\bf k''}\beta}(t'')\rangle\k_\beta
\,dt'dt'' \!\!\!&=&\!\!\! J_{0k''}t 
\left(\frac{1+{\bar\Omega}''/\Omega_{\rm rd}''}{1+2\Omega''/\Omega''_{\rm rd}}\right)^2
\frac{k'(k'-k\cos\theta)\sin\theta\cos\varphi}{k''^2},
\\
\int_0^t\!\!\int_0^t
\k_\alpha\langle\ViT_{{\bf k''}\alpha}(t')\ViT^*_{{\bf k''}\beta}(t'')\rangle\k_\beta\,dt'dt'' 
\!\!\!&=&\!\!\! J_{0k''}t\left(1+{\bar\Omega}''/\Omega_{\rm rd}''\right)^2
\frac{k'^2\sin^2\theta}{(k'-k\cos\theta)^2+k^2\sin^2\theta\sin^2\varphi}
\nonumber\\
{}\!\!\!&\times&\!\!\!\!\!
\left[\sin^2\varphi + \left(1+2\Omega''/\Omega''_{\rm rd}\right)^{-2}
\frac{(k'-k\cos\theta)^2\cos^2\varphi}{k''^2}\right]\!.
\qquad\quad
\label{THIRD_DOUBLE_TIME_INTEGRAL}
\eeq
Here, ${\bar\Omega}''$, $\Omega''$ and $\Omega''_{\rm rd}$ depend on $k''$ and 
on $\mu''^2$ [see eqs.~(\ref{OMEGA}),~(\ref{OMEGA_DAMP})]. In turn, $k''$ and 
$\mu''$ are functions of $k$, $k'$, $\theta$ and $\varphi$ [see 
eqs.~(\ref{kk'k''}),~(\ref{MU''}). Now, we substitute formulas~(\ref{MU_dk}) 
and~(\ref{FIRST_DOUBLE_TIME_INTEGRAL})--(\ref{THIRD_DOUBLE_TIME_INTEGRAL}) into 
equations~(\ref{T})--(\ref{T'''}). The factors, which we obtain in these 
equations after integration over $d^2{\bf\k}=\sin\theta\;d\theta\:d\varphi$, 
depend only on $k$ and $k'$, so they can be exchanged with the integrations 
over $d^2{\bf\k'}$. Combining the results together in 
equation~(\ref{BiTS_INTEGRATED_FOUR_TERMS}), we finally obtain 
equations~(\ref{BiTS_TERM})--(\ref{OMEGA_RATIO_FIRST}).
\clearpage

\clearpage

\section{The derivation of equation~(\ref{F_M_INTEGRAL})}
\label{SMALL_SCALES_DERIVATION}

First, we use formula $d^3{\bf k}=dk\,d^2{\bf\k}$, 
equations~(\ref{BiTS_INTEGRATED_FOUR_TERMS})--(\ref{T'''}) and 
equation~(\ref{BiiT_AVERAGED}) to obtain the term in the brackets $\{...\}$
in equation~(\ref{F_M_INTEGRAL}),
\beq
\{...\} \!\!&=&\!\! \int_0^\infty\!F(k)\, dk \int\! k^2\,
\langle|\BiTbf_{\bf k}(t)|^2\rangle\,d^2{\bf\k}+\int_0^\infty\!F(k)\,dk \int\! k^2\,
\Big[\langle\BiiT_{{\bf k}\alpha}(t)\rangle\BoT^*_{{\bf k}\alpha}+{\rm c.c.}\Big]\,d^2{\bf\k}
\nonumber\\
{}\!\!&=&\!\! {\left(\frac{L}{2\pi}\right)}^{\!3}
\Big[{\cal T}_\diamond-{\cal T}'_\diamond-{\cal T}''_\diamond+{\cal T}'''_\diamond
+{\cal T}^{\rm iv}_\diamond\Big],
\qquad
\label{F_B_INTEGRAL}
\\
{\cal T}_\diamond \!\!&=&\!\! \int_{-\infty}^\infty\int_{-\infty}^\infty
\mu^2 k^2 F(k)\,|\BoT_{\bf k'}|^2 \,d^3{\bf k'}d^3{\bf k}
\int_0^t\!\!\int_0^t 
\langle\ViT_{{\bf k''}\alpha}(t')\ViT^*_{{\bf k''}\alpha}(t'')\rangle
\,dt' dt''
\nonumber\\
{}\!\!&=&\!\!
\int_{-\infty}^\infty \!\!|\BoT_{\bf k'}|^2 d^3{\bf k'} \!
\int_{-\infty}^\infty \!\!\mu''^2 k''^2 F(k)\,d^3{\bf k''} 
\int_0^t\!\!\int_0^t 
\langle\ViT_{{\bf k''}\alpha}(t')\ViT^*_{{\bf k''}\alpha}(t'')\rangle
\,dt' dt'',
\quad
\label{T_d}
\\
{\cal T}'_\diamond  \!\!&=&\!\!
\int_{-\infty}^\infty\int_{-\infty}^\infty
\mu k F(k)\,|\BoT_{\bf k'}|^2 \,d^3{\bf k'}d^3{\bf k} 
\int_0^t\!\!\int_0^t 
\bo_\alpha\langle\ViT_{{\bf k''}\alpha}(t')\ViT^*_{{\bf k''}\beta}(t'')\rangle k_\beta
\,dt' dt''
\nonumber\\
{}\!\!&=&\!\!
\int_{-\infty}^\infty \!\!|\BoT_{\bf k'}|^2 d^3{\bf k'} \!
\int_{-\infty}^\infty \!\!\mu''k'' F(k)\, d^3{\bf k''}
\int_0^t\!\!\int_0^t 
\bo_\alpha\langle\ViT_{{\bf k''}\alpha}(t')\ViT^*_{{\bf k''}\beta}(t'')\rangle k'_\beta
\,dt' dt'',
\qquad
\label{T'_d}
\\
{\cal T}''_\diamond  \!\!&=&\!\!  
{\cal T}'_\diamond,
\label{T''_d}
\\
{\cal T}'''_\diamond  \!\!&=&\!\!
\int_{-\infty}^\infty\int_{-\infty}^\infty
F(k)\,|\BoT_{\bf k'}|^2 \,d^3{\bf k'}d^3{\bf k}
\int_0^t\!\!\int_0^t 
k_\alpha\langle\ViT_{{\bf k''}\alpha}(t')\ViT^*_{{\bf k''}\beta}(t'')\rangle k_\beta
\,dt' dt''
\nonumber\\
{}\!\!&=&\!\!
\int_{-\infty}^\infty \!\!|\BoT_{\bf k'}|^2 d^3{\bf k'} \!
\int_{-\infty}^\infty \!\! F(k)\,d^3{\bf k''}
\int_0^t\!\!\int_0^t 
k'_\alpha\langle\ViT_{{\bf k''}\alpha}(t')\ViT^*_{{\bf k''}\beta}(t'')\rangle k'_\beta
\,dt' dt'',
\quad
\label{T'''_d}
\\
{\cal T}^{\rm iv}_\diamond \!\!&=&\!\!
{}-2 \!\int_{-\infty}^\infty\!\!
F(k)\,|\BoT_{\bf k}|^2 \,d^3{\bf k}
\!\int_{-\infty}^\infty\!d^3{\bf k''}
\int_0^t\!dt'\!\int_0^{t'}\!dt''\;
k_\alpha\langle\ViT_{-{\bf k''}\alpha}(t'')\,\ViT_{{\bf k''}\beta}(t')\rangle k_\beta
\nonumber\\
{}\!\!&=&\!\!
{}- \int_{-\infty}^\infty\!\!
F(k)\,|\BoT_{\bf k}|^2 \,d^3{\bf k}
\!\int_{-\infty}^\infty\!d^3{\bf k''}
\int_0^t\!\!\int_0^t 
k_\alpha\langle\ViT_{-{\bf k''}\alpha}(t'')\,\ViT_{{\bf k''}\beta}(t')\rangle k_\beta
\,dt' dt''.
\qquad
\label{Tiv_d}
\eeq
Here, to obtain the final results in equations~(\ref{T_d})--(\ref{T'''_d}), 
we change the integration over ${\bf k}$ in these equations to integration 
over ${\bf k''}$, using ${\bf k}={\bf k'}+{\bf k''}$. We use 
$k''_\alpha\ViT_{{\bf k''}\alpha}$, and therefore, 
$k_\alpha\ViT_{{\bf k''}\alpha}=k'_\alpha\ViT_{{\bf k''}\alpha}$. We also 
use $\mu k=\bobf\cdot{\bf k}=\bobf\cdot{\bf k''}=k''(\bobf\cdot{\bf\k''})
=\mu''k''=k''\sin\theta\cos\varphi$ (see Fig.~\ref{FIGURE_bo_and_k}B). 
To obtain the final result in equation~(\ref{Tiv_d}), we use 
equations~(\ref{ViT_ViT_AVERAGED_NEW}) and~(\ref{H_F_INTEGRAL_TT}).
Next, we calculate ${\cal T}_\diamond$, ${\cal T}'_\diamond$ and ${\cal T}'''_\diamond$
separately, up to the second order in $k''\ll k,k'$. Refer to 
Figure~\ref{FIGURE_bo_and_k}B.

First, following Kulsrud and Anderson (1992), in equations~(\ref{T_d})--(\ref{T'''_d}) 
we expand $F(k)$ in $k''\ll k$ at point $k'$ up to the second order. We have, see 
Figure~\ref{FIGURE_bo_and_k}B,
\beq
k \!\!&=&\!\! k'+k''\cos\theta+(k''^2/2k')\sin^2\theta\,,
\qquad
\label{K_EXPANSION}
\\
F(k) \!\!&=&\!\! F(k') + \frac{dF}{dk'}k''\cos\theta +\frac{1}{2k'}\frac{dF}{dk'}k''^2\sin^2\theta
+\frac{1}{2}\frac{d^2F}{dk'^2}k''^2\cos^2\theta,
\label{F_EXPANSION}
\eeq
(Kulsrud \& Anderson 1992; Malyshkin 2001).

Second, we calculate ${\cal T}_\diamond$, given by equation~(\ref{T_d}).
Because $\mu''^2k''^2$ is of the second order in $k''$, we need to keep only the zero 
order term in expansion~(\ref{F_EXPANSION}) for $F(k)$. Thus, we have
\beq
{\cal T}_\diamond \!\!&=&\!\!
\int_{-\infty}^\infty \!F(k')\,|\BoT_{\bf k'}|^2 \,d^3{\bf k'}
\int_{-\infty}^\infty \!\mu''^2k''^2 \,d^3{\bf k''}\int_0^t \!\!\int_0^t 
\langle\ViT_{{\bf k''}\alpha}(t')\ViT^*_{{\bf k''}\alpha}(t'')\rangle\,dt' dt'' 
\nonumber\\
\!\!&=&\!\!
t \int_0^\infty \!F(k')\,dk' \int\! k'^2\,|\BoT_{\bf k'}|^2 \,d^2{\bf\k'}
\int_0^\infty k''^4 J_{0k''}\,dk'' \int_0^\pi\! d\theta\, \sin^3\theta
\nonumber\\
\!\!&&\!\!{}\times
\int_0^{2\pi}\! d\varphi\, \cos^2\varphi
\left(1+{\bar\Omega}''/\Omega''_{\rm rd}\right)^2
\left[1+\left(1+2\Omega''/\Omega''_{\rm rd}\right)^{-2}\right],
\label{T_d_RESULT}
\eeq
where we use $d^3{\bf k''} = k''^2dk''\sin\theta\,d\theta\,d\varphi$,
$\mu''=\sin\theta\cos\varphi$ and equation~(\ref{FIRST_DOUBLE_TIME_INTEGRAL}). 
Here and below, functions ${\bar\Omega}''$, $\Omega''$ and $\Omega_{\rm rd}''$ 
depend on $k''$ and $\mu''^2=\sin^2\theta\cos^2\varphi$, see 
equations~(\ref{OMEGA}) and~(\ref{OMEGA_DAMP}).

Third, we calculate ${\cal T}'_\diamond$, given by equation~(\ref{T'_d}). 
Because $\mu''k''$ is of the first order in $k''$, we need to keep only the zero and
the first order terms in expansion~(\ref{F_EXPANSION}) for $F(k)$. We have
\beq
{\cal T}'_\diamond \!\!&=&\!\! \int_{-\infty}^\infty \!\!F(k')|\BoT_{\bf k'}|^2 d^3{\bf k'} 
\! \int_{-\infty}^\infty \!\!\mu''k'' d^3{\bf k''} \int_0^t\!\!\int_0^t 
\bo_\alpha\langle\ViT_{{\bf k''}\alpha}(t')\ViT^*_{{\bf k''}\beta}(t'')\rangle k'_\beta
\,dt' dt''
\nonumber\\
{}\!\!&+&\!\!
\int_{-\infty}^\infty \!\frac{dF}{dk'}|\BoT_{\bf k'}|^2 d^3{\bf k'}\!\!
\int_{-\infty}^\infty \!\!\!\mu''k''^2 \cos\theta\, d^3{\bf k''}\!\!\int_0^t\!\!\int_0^t 
\!\bo_\alpha\langle\ViT_{{\bf k''}\alpha}(t')\ViT^*_{{\bf k''}\beta}(t'')\rangle k'_\beta
\,dt' dt''
\nonumber\\
\!\!&=&\!\!
-t \int_0^\infty k'\,\frac{dF}{dk'}\,dk'\int\!k'^2\,|\BoT_{\bf k'}|^2\,d^2{\bf\k'}
\int_0^\infty k''^4J_{0k''}\,dk''\int_0^\pi d\theta\, \sin^3\theta\cos^2\theta
\nonumber\\
\!\!&&\!\!\quad{}\times \int_0^{2\pi} d\varphi\, \cos^2\varphi
\left(1+{\bar\Omega}''/\Omega''_{\rm rd}\right)^2
\left(1+2\Omega''/\Omega''_{\rm rd}\right)^{-2}.
\label{T'_d_RESULT}
\eeq
Here, we use equations~(\ref{ViT_ViT_AVERAGED_NEW}),~(\ref{H_F_INTEGRAL_TT}),
$k'_\alpha\bo_\alpha=0$, $d^3{\bf k''} = k''^2dk''\sin\theta\,d\theta\,d\varphi$, 
$\mu''=\sin\theta\cos\varphi$. The first term in the first line of 
equation~(\ref{T'_d_RESULT}) vanishes after the integration over $\theta$ because 
the integrand is an odd function of $\cos\theta$. 

Fourth, we calculate ${\cal T}'''_\diamond$, given by equation~(\ref{T'''_d}),
keeping all terms in expansion~(\ref{F_EXPANSION}) for $F(k)$,
\beq
{\cal T}'''_\diamond  \!\!&=&\!\! \int_{-\infty}^\infty \!\!F(k')
|\BoT_{\bf k'}|^2 d^3{\bf k'} \! \int_{-\infty}^\infty \!\! d^3{\bf k''} \int_0^t\!\!\int_0^t 
k'_\alpha\langle\ViT_{{\bf k''}\alpha}(t')\ViT^*_{{\bf k''}\beta}(t'')\rangle k'_\beta
\,dt' dt''
\nonumber\\
{}\!\!&+&\!\! \int_{-\infty}^\infty \frac{dF}{dk'} |\BoT_{\bf k'}|^2 d^3{\bf k'} \!
\int_{-\infty}^\infty \!\! k''\cos\theta\,d^3{\bf k''} \int_0^t\!\!\int_0^t 
k'_\alpha\langle\ViT_{{\bf k''}\alpha}(t')\ViT^*_{{\bf k''}\beta}(t'')\rangle k'_\beta
\,dt' dt''
\nonumber\\
{}\!\!&+&\!\! \int_{-\infty}^\infty \frac{1}{2k'}\frac{dF}{dk'} |\BoT_{\bf k'}|^2 d^3{\bf k'} \!
\int_{-\infty}^\infty \!\! k''^2\sin^2\theta\,d^3{\bf k''} \int_0^t\!\!\int_0^t 
k'_\alpha\langle\ViT_{{\bf k''}\alpha}(t')\ViT^*_{{\bf k''}\beta}(t'')\rangle k'_\beta
\,dt' dt''
\nonumber\\
{}\!\!&+&\!\!\! \int_{-\infty}^\infty \!\frac{1}{2}\frac{d^2F}{dk'^2}
|\BoT_{\bf k'}|^2 d^3{\bf k'} \!\! 
\int_{-\infty}^\infty \!\! k''^2\cos^2\theta\,d^3{\bf k''}\!\int_0^t\!\!\int_0^t \!
k'_\alpha\langle\ViT_{{\bf k''}\alpha}(t')\ViT^*_{{\bf k''}\beta}(t'')\rangle k'_\beta
\,dt' dt''
\nonumber\\
\!\!&=&\!\! {}-{\cal T}^{\rm iv}_\diamond
\nonumber\\
{}\!\!&+&\!\! \frac{t}{2}\int_0^\infty\! k'\,\frac{dF}{dk'}\,dk' \int\! k'^2\, 
|\BoT_{\bf k'}|^2 d^2{\bf\k'}\! \int_0^\infty k''^4J_{0k''}\,dk'' 
\int_0^\pi d\theta\, \sin^5\theta
\nonumber\\
\!\!&&\!\! \;{}\times \int_0^{2\pi} d\varphi\,
\left(1+{\bar\Omega}''/\Omega''_{\rm rd}\right)^2
\left\{ 1-\left[1-\left(1+2\Omega''/\Omega''_{\rm rd}\right)^{-2}\right] 
\frac{\cos^2\theta\cos^2\varphi}{\cos^2\theta\cos^2\varphi+\sin^2\varphi}
\right\}
\nonumber\\
{}\!\!&+&\!\! \frac{t}{2}\int_0^\infty\! k'^2\,\frac{d^2F}{dk'^2}\,dk'
\int\! k'^2\,|\BoT_{\bf k'}|^2 d^2{\bf\k'}\! \int_0^\infty k''^4J_{0k''}\,dk''
\int_0^\pi d\theta\, \sin^3\theta\cos^2\theta
\nonumber\\
\!\!&&\!\! \;{}\times \int_0^{2\pi} d\varphi\, 
\left(1+{\bar\Omega}''/\Omega''_{\rm rd}\right)^2
\left\{ 1-\left[1-\left(1+2\Omega''/\Omega''_{\rm rd}\right)^{-2}\right] 
\frac{\cos^2\theta\cos^2\varphi}{\cos^2\theta\cos^2\varphi+\sin^2\varphi}\right\}.
\qquad
\label{T'''_d_RESULT}
\eeq
Here, the first term in the first line is equal to minus ${\cal T}^{\rm iv}_\diamond$, 
given by equation~(\ref{Tiv_d}). To calculate the other three terms (on the 2nd, 3rd and 
4th lines), we again use equations~(\ref{ViT_ViT_AVERAGED_NEW}),~(\ref{H_F_INTEGRAL_TT}), 
$k'_\alpha\bo_\alpha=0$, $d^3{\bf k''} = k''^2dk''\sin\theta\,d\theta\,d\varphi$, 
$\mu''=\sin\theta\cos\varphi$. The term in the second line of 
equation~(\ref{T'''_d_RESULT}) vanishes after the integration over $\theta$ 
because the integrand is an odd function of $\cos\theta$.

Next, we substitute equations~(\ref{T''_d}),~(\ref{T_d_RESULT}),~(\ref{T'_d_RESULT})
and (\ref{T'''_d_RESULT}) into equation~(\ref{F_B_INTEGRAL}). The 
$-{\cal T}^{\rm iv}_\diamond$ term of equation~(\ref{T'''_d_RESULT}) cancels the 
${\cal T}^{\rm iv}_\diamond$ term in equation~(\ref{F_B_INTEGRAL}). Then, we 
substitute the result, obtained in equation~(\ref{F_B_INTEGRAL}), into the second 
line of equation~(\ref{F_M_INTEGRAL}) and, using equation~(\ref{M_K}), we find
\beq
\int_0^\infty\!F(k)\,\frac{\partial M}{\partial t}\,dk \!\!&=&\!\!
{\left(\frac{L}{2\pi}\right)}^{\!3}\! \int_0^\infty \left[\lambda_0F(k')
+\lambda_1k'\frac{dF}{dk'} +\lambda_2k'^2\frac{d^2F}{dk'^2}\right] M(0,k')\,dk',
\qquad\quad
\label{INTEGRAL_SMALL_SCALES_FIRST}
\eeq
\beq
\lambda_0 \!\!&=&\!\! \int_0^\infty k''^4 J_{0k''}\,dk''
\int_0^\pi\! d\theta\, \sin^3\theta \int_0^{2\pi}\! d\varphi\, \cos^2\varphi
\left(1+{\bar\Omega}''/\Omega''_{\rm rd}\right)^2
\left[1+\left(1+2\Omega''/\Omega''_{\rm rd}\right)^{-2}\right],
\qquad
\label{lambda_0}
\\
\lambda_1 \!\!&=&\!\! \int_0^\infty k''^4 J_{0k''}\,dk''
\int_0^\pi\! d\theta\, \sin^3\theta \int_0^{2\pi}\! d\varphi 
\left(1+{\bar\Omega}''/\Omega''_{\rm rd}\right)^2
\left\{ 2\cos^2\theta\cos^2\varphi+\frac{1}{2}\sin^2\theta
\right.
\nonumber\\
\!\!&&\!\!\left. {}-\left[1-\left(1+2\Omega''/\Omega''_{\rm rd}\right)^{-2}\right]
\left(2+\frac{1}{2}\,\frac{\sin^2\theta}{\cos^2\theta\cos^2\varphi+\sin^2\varphi}\right)
\cos^2\theta\cos^2\varphi\right\}\!,
\label{lambda_1}
\\
\lambda_2 \!\!&=&\!\! \frac{1}{2}\int_0^\infty k''^4J_{0k''}\,dk''
\int_0^\pi d\theta\, \sin^3\theta\cos^2\theta
\nonumber\\
\!\!&&\!\! {}\times \int_0^{2\pi} d\varphi\, 
\left(1+{\bar\Omega}''/\Omega''_{\rm rd}\right)^2\!
\left\{ 1-\left[1-\left(1+2\Omega''/\Omega''_{\rm rd}\right)^{-2}\right] 
\frac{\cos^2\theta\cos^2\varphi}{\cos^2\theta\cos^2\varphi+\sin^2\varphi}\right\}\!.
\quad
\label{lambda_2}
\eeq
Now, we integrate the right-hand-side of equation~(\ref{INTEGRAL_SMALL_SCALES_FIRST}) 
by parts over some extent in $k'$ and choose $F(k')$, so that it and its derivative 
$dF/dk'$ vanish at the end points. We finally obtain the third line of 
equation~(\ref{F_M_INTEGRAL}), by dropping the double primes, $''$, in 
equations~(\ref{lambda_0})--(\ref{lambda_2}), and introducing the new 
constants $\Gamma$, $\Lambda_1$ and $\Lambda_0$, given by 
equations~(\ref{CAPITAL_GAMMA})--(\ref{LAMBDA_0}).

\end{document}